\documentclass[preprint]{aastex}
\usepackage{graphicx}        % For eps figures, newer & more powerfull
\usepackage{natbib}         % For citations: redefine \cite commands
\usepackage{amssymb}        % useful mathemati cal symbols
\usepackage{amsmath}
\usepackage{color}           % For color text: \color commandc
\usepackage{url}             % For breaking URLs easily trough lines
\usepackage{epstopdf}
\usepackage{soul}
\newcommand\bu{{\boldsymbol{u}}}
\newcommand\bU{{\boldsymbol{U}}}
\newcommand\be{{\boldsymbol{e}}}

\newcommand\bk{{\boldsymbol{k}}}

\renewcommand\Re{{\rm Re}}
\newcommand\Ri{{\rm Ri}}
\newcommand\Pe{{\rm Pe}}
\renewcommand\Pr{{\rm Pr}}
\newcommand\Ra{{\rm Ra}}

\begin{document}

\pagestyle{empty} %No headings for the first pages.

%% Title Page %%%%%%%%%%%%%%%%%%%%%%%%%%%%%%%%%%%%%%%%%%%%%%%
%% ==> Write your text here or include other files.

%% The simple version:
\title{Turbulent transport by diffusive stratified shear flows: from local to global models. Part I: Numerical simulations of a stratified plane Couette flow}

\author{Pascale Garaud, \\
Department of Applied Mathematics and Statistics, Baskin School of Engineering, \\University of California at Santa Cruz, 1156 High Street, Santa Cruz CA 95064.\\
Damien Gagnier, \\
Institut de Recherche en Astrophysique et Plan\'etologie (IRAP), 14, avenue Edouard Belin, 31400 Toulouse, France.\\  
Jan Verhoeven,\\
Department of Earth and Planetary Sciences, \\University of California at Santa Cruz, 1156 High Street, Santa Cruz CA 95064.
 }

%\date{} %%If commented, the current date is used.
\maketitle

% The nice version:
%\input{titlepage} %%You need a file 'titlepage.tex' for this.
%% ==> TeXnicCenter supplies a possible titlepage file
%% ==> with its templates (File | New from Template...).

%\begin{abstract}
\vspace{1cm}
\centerline{\bf Abstract} 
Shear-induced turbulence could play a significant role in mixing momentum and chemical species in stellar radiation zones, as discussed by Zahn (1974). In this paper we analyze the results of direct numerical simulations of stratified plane Couette flows, in the limit of rapid thermal diffusion, to measure the turbulent viscosity and the turbulent diffusivity of a passive tracer as a function of the local shear and the local stratification. We find that the stability criterion proposed by Zahn (1974), namely that the product of the gradient Richardson number and the Prandtl number must be smaller than a critical values $(J\Pr)_c$ for instability, adequately accounts for the transition to turbulence in the flow, with $(J\Pr)_c \simeq 0.007$. This result recovers and confirms the prior findings of Prat et al. (2016). Zahn's model for the turbulent diffusivity and viscosity (Zahn 1992), namely that the mixing coefficient should be proportional to the ratio of the thermal diffusivity to the gradient Richardson number, does not satisfactorily match our numerical data. It fails (as expected) in the limit of large stratification where the Richardson number exceeds the aforementioned threshold for instability, but it also fails in the limit of low stratification where the turbulent eddy scale becomes limited by the computational domain size. We propose a revised model for turbulent mixing by diffusive stratified shear instabilities, that now properly accounts for both limits, fits our data satisfactorily, and recovers Zahn's 1992 model in the limit of large Reynolds numbers.

\section{Introduction}
\label{sec:intro}

Thanks to recent advances in supercomputing it is now possible to run numerical experiments designed to quantify the rates of turbulent mixing of selected quantities (heat, angular momentum, composition, etc) in the presence of various hydrodynamic and magnetohydrodynamic instabilities, in parameter regimes appropriate of astrophysical plasmas that cannot be achieved in more traditional laboratory experiments. This raises the engaging prospect of finally being able to constrain models of non-canonical mixing in stars through first-principles theory and experiments rather than through observations alone. 

A strong potential candidate for driving vertical mixing in stellar radiation zones is the shear instability, which has gained significant popularity since the discovery of relatively strong radial shear layers via helio- and asteroseismology, in the Sun \citep{JCDSchou88,Brownal89} and in Red Giant Branch (RGB) stars \citep{Deheuvelsetal12,Deheuvelsetal14}. However, the radiation zones of these stars are usually too strongly stratified to be shear-unstable, at least if thermal diffusion is ignored. Indeed, the standard criterion for instability of inviscid, non-diffusive stratified shear flows to infinitesimal perturbations is the Richardson criterion \citep[formally proved by][]{Miles61,Howard61}, which states that the local gradient Richardson number 
\begin{equation}
J=\frac{N^2}{S^2}\, ,
\end{equation}
where $N$ is the Brunt-V\"ais\"al\"a frequency and $S$ is the local shearing rate, must drop below 1/4 somewhere in the flow for instability. It is sometimes argued that finite amplitude non-diffusive instabilities may exist for somewhat larger values of $J$, but energetic arguments \citep{Richardson1920} suggest that $J$ cannot be much larger than one for sustained turbulent motions to exist. 

In thermally-stratified systems, \citet{Townsend58} and \citet{Dudis1974} noted, however, that thermal radiation and/or thermal diffusion lets the perturbed fluid adjust thermally to its surroundings, thereby reducing its stabilizing buoyancy excess or deficit compared with the background as it moves up and down. In an optically thick fluid, the ratio of the thermal diffusion timescale to the shearing timescale  for a shear layer of vertical lengthscale $L$ is measured by the global P\'eclet number, 
\begin{equation}
{\rm Pe}_L = \frac{S L^2}{\kappa_T}\, ,
\label{eq:PeL}
\end{equation}
where $\kappa_T$ is the thermal diffusivity. Linear stability analyses have shown that having ${\rm Pe}_L \ll 1$ can destabilize stratified shear layers that would otherwise be stable in the non-diffusive case \citep{Jones1977,Lignieresetal1999,Garaudal15}, and typically raises the critical Richardson number for instability by a factor proportional to ${\rm Pe}_L^{-1}$. Note that unfortunately there is no simple formal criterion for linear instability equivalent to the Miles-Howard theorem when thermal diffusion is accounted for, even in the inviscid limit. 
%In fact, anecdotal examples exist of strongly diffusive stratified shear flows profiles that are unstable (in the inviscid limit, and in infinitely large domains) regardless of $J$, such as the hyperbolic tangent shear layer of \citet{Lignieresetal1999}, while other profiles under the same conditions are only linearly unstable if the product of the Richardson number and the P\'eclet number, $J {\rm Pe}_L$, drops below a certain critical value, such as the sinusoidal profiles studied by \citet{Garaudal15}. Finally, to add to the confusion perhaps, shear flows that have constant $S$ and constant $N$ are always linearly stable \citep{Knobloch84} -- whether diffusive or not. 

In any case, linear theory is of limited use, since it neither addresses the possibility of destabilization by finite-amplitude perturbations, nor the question of the saturation of the instability, which is necessary to understand the properties of the turbulent flows that later develop. In an attempt to go beyond linear theory, Zahn therefore proposed a now widely-used two-part model for shear instabilities in stars that involves a revised criterion for instability \citep{Zahn1974} and a turbulent mixing prescription \citep{Zahn92}. We describe both in turn. 

\subsection{Stability criterion}
\label{sec:introstab}

\begin{figure}[h]
  \centerline{\includegraphics[width=0.5\textwidth]{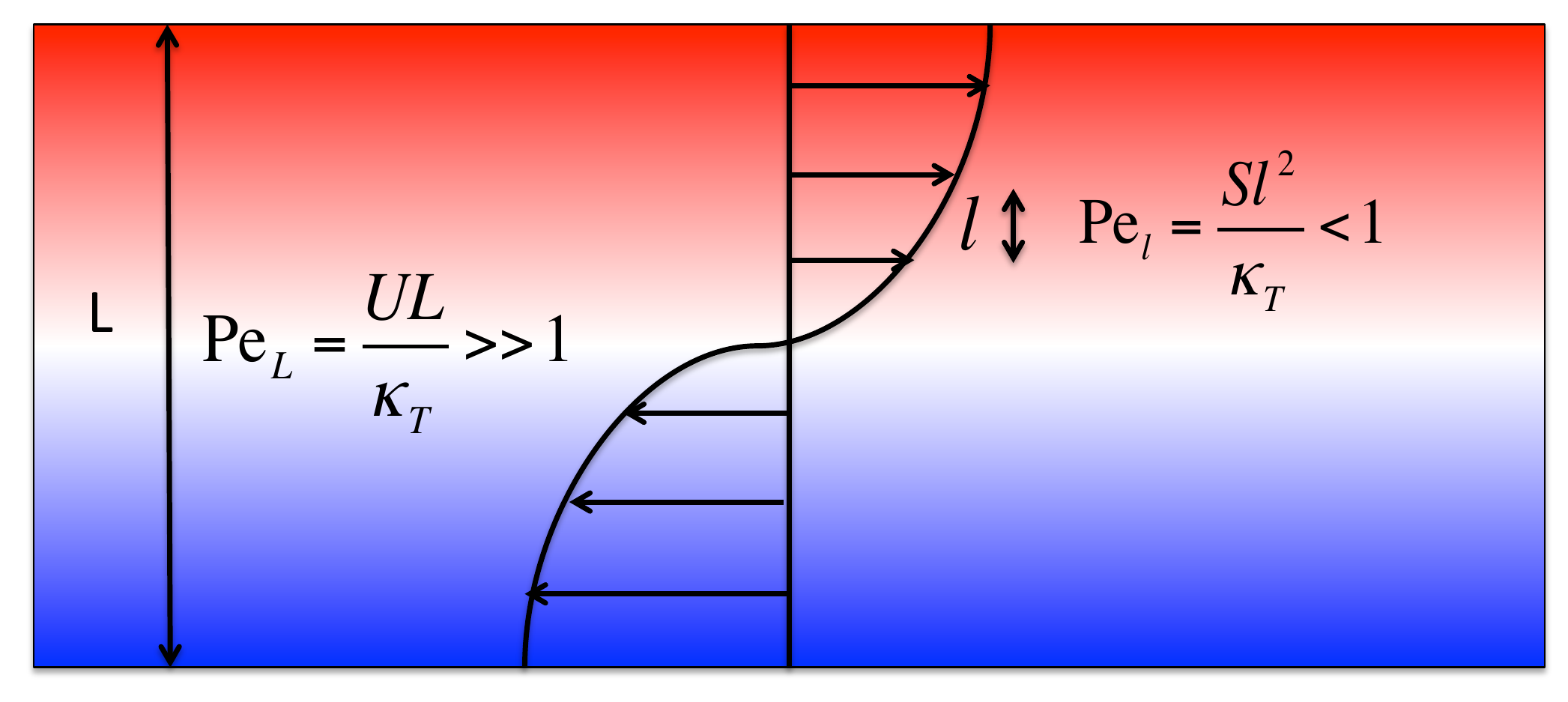}}
  \caption{Illustration of the global vs. local properties of stratified shear flows. The shading represents the stratification, from denser fluid (blue) to lighter fluid (red). On the global scale, the shear may not be constant but will have a well-defined amplitude $U$ and lengthscale $L$, defining a global P\'eclet number $\Pe_L = UL/\kappa_T$ which is likely much larger than one. On small enough scales $l$, however, one can define an eddy P\'eclet number $\Pe_l = Sl^2/\kappa_T$ that is much smaller than one. On these small scales, the local shearing rate $S$ is approximately constant. Zahn's models assume that the turbulence develop on these small vertical scales, so that both stability criterion and mixing can be treated in a local fashion. }
\label{fig:modelillustration}
\end{figure}

To address the question of stability, \citet{Zahn1974} noted that, allowing for finite amplitude instabilities, structures of height $l$ much smaller than the shear scaleheight $L$ can in principle be destabilized provided their eddy P\'eclet number $ {\rm Pe}_l \equiv S l^2/\kappa_T$ is much smaller than 1 (even if  ${\rm Pe}_L   \gg 1$), see Figure \ref{fig:modelillustration}. By analogy with the results of \citet{Townsend58} and \citet{Dudis1974}, he proposed that instability on the scale $l$ may occur, in the inviscid limit, as long as  
\begin{equation}
J {\rm Pe}_l < (J{\rm Pe})_c\, ,
\label{eq:crit1}
\end{equation}
where $(J{\rm Pe})_c$ is a constant of order one. 

Accounting for the effect of viscosity limits the range of scales that can be destabilized this way, requiring in addition that 
\begin{equation}
{\rm Re}_l \equiv \frac{S l^2 }{ \nu} > {\rm Re}_c\, ,
\label{eq:Recrit}
\end{equation}
where $\nu$ is the viscosity. The constant ${\rm Re}_c$ is a critical Reynolds number for finite-amplitude instability, which he picks to be of order $10^3$. Combining these two criteria leads to Zahn's local criterion for diffusive stratified shear instabillities \citep{Zahn1974}, namely that a flow is unstable provided
\begin{equation}
J \Pr = J \frac{{\rm Pe}_l }{ {\rm Re}_l} <  \frac{(J{\rm Pe})_c }{{\rm Re}_c} \equiv (J \Pr)_c
\label{eq:crit2}
\end{equation}
where $\Pr = \nu/\kappa_T$ is the Prandtl number. This defines the critical Richardson-Prandtl number for shear instabilities $(J \Pr)_c$ which, according to the arguments given above \citep{Zahn1974}, should be of order $10^{-3}$. Note that, by contrast with (\ref{eq:crit1}), (\ref{eq:crit2}) is now independent of the turbulent eddy scale, which is quite convenient since the latter is not known a priori. 

While Zahn's criterion is based on somewhat heuristic arguments, two important pieces of formal theory support it. First, 
as established by \citet{Lignieres1999}, the dynamics of strongly diffusive stratified shear instabilities, that is, in the limit of asymptotically low P\'eclet numbers, can {\it only} depend on two non-dimensional parameters (aside from possible geometric effects), namely the global Reynolds number (${\rm Re}_L \equiv \frac{S L^2}{\nu}$) and the {\it product} $J {\rm Pe}_L$. Hence any stability criterion, should it exist, must be expressed in terms of ${\rm Re}_L$ and $J {\rm Pe}_L$, rather than ${\rm Re}_L$, $J$ and ${\rm Pe}_L$ separately. It is easy to verify that it is indeed true in Zahn's criterion since $J {\rm Pr} = J {\rm Pe}_L / {\rm Re}_L$. 

Secondly, in this low P\'eclet number limit \citep{Lignieres1999} it is possible to obtain formal results on the energy stability of stratified shear flows. Energy stability is arguably a more useful concept than linear stability because it implies that any initial perturbation must decay, regardless of their amplitude or form, while linear stability only implies that infinitesimal perturbations must decay. As a result, establishing the energy stability of a shear layer guarantees that there can be no shear-induced turbulent mixing (unless other physics are added that could further destabilize it). For instance, \citet{Bischoff13} showed that a low P\'eclet number linear shear in a linearly stratified background (i.e. where $S$ and $N$ are both constant) is energy stable provided $J {\rm Pr} > 1/4$. A similar energy stability criterion was obtained by \citet{Garaudal15} for any periodic shear flow, albeit with a constant that is different from 1/4 and that depends on the shape of the domain considered. One may therefore postulate that a reasonable general {\it energy stability} criterion for diffusive stratified shear layers in stellar interiors is 
\begin{equation}
J \Pr >  (J \Pr)_e\, ,
\label{eq:escrit}
\end{equation}
where $(J \Pr)_e$ is a constant of order unity that may depend on the aspect ratio of the domain. The scaling of this criterion with $J$ and ${\rm Pr}$ is identical to the one proposed by \citet{Zahn1974}, suggesting that his physically-motivated derivation could be a plausible explanation for the more formal but significantly less intuitive energy stability criterion of \citet{Bischoff13} and \citet{Garaudal15}.

Despite these supporting arguments, however, neither linear theory nor energy stability theory can really address the question of whether Zahn's criterion is indeed a good predictor for the transition to turbulence in stratified shear flows. A linearly stable shear flow may still be the seat of vigorous turbulent mixing. Meanwhile, a shear layer being outside of the energy-stable region of parameter space does not necessarily imply turbulence. What really matters in terms of stellar mixing is neither energy stability nor linear instability, but instead the more elusive threshold for sustained turbulence driven by finite amplitude instabilities. This threshold can only be obtained using 3D direct numerical simulations (DNS).

\subsection{Mixing model}
\label{sec:intromix}

Later, Zahn expanded his theory for stratified shear instabilities in stellar interiors, presenting a model for the induced turbulent diffusivity $D_{\rm turb}$ \citep{Zahn92}. 
Starting from dimensional analysis, he first proposed that 
\begin{equation}
D_{\rm turb} = \beta S l_e^2\, ,
\label{eq:dturblz}
\end{equation}
where $l_e$ is the lengthscale of the dominant turbulent eddies and $\beta$ is a constant of order unity. He then argued that $l_e$ can be estimated by assuming it is also the scale of turbulent eddies that are just marginally unstable according to criterion (\ref{eq:crit1}), or in other words, the largest lengthscale for which (\ref{eq:crit1}) holds: 
\begin{equation}
J \frac{ S l_e^2}{\kappa_T} =  (J \Pe)_c \, ,
\end{equation}
which implies 
\begin{equation}
l_e = \sqrt{ \frac{(J \Pe)_c \kappa_T}{JS}} = \sqrt{ \frac{(J \Pe)_c }{J \Pe_L }} L  \equiv l_{\rm Z} \, .
\label{eq:zahnscale}
\end{equation}
We shall henceforth refer to this scale as the Zahn scale, and denote it as $l_{\rm Z}$. Using $l_{\rm Z}$ in $D_{\rm turb}$, we arrive at
\begin{equation}
D_{\rm turb} = \beta (J \Pe)_c  \frac{\kappa_T}{J} \equiv C \frac{\kappa_T}{J} \mbox{   where   } C = \beta (J \Pe)_c 
\label{eq:zahndt}
\end{equation}
is a universal constant. 

Two points are worth noting. First, this model does not specify what quantity $D_{\rm turb}$ actually mixes, and in this sense, could be applied both to the transport of chemical species or of momentum, albeit perhaps with a different constant $\beta$. Secondly, this model must clearly fail both in the limit of very large stratification (i.e when the system is stabilized for $J{\rm Pr}>(J{\rm Pr})_c$ so $D_{\rm turb}$ must drop to zero) and in the limit of very low stratification (where $J$ tends to 0 and $D_{\rm turb}$ would formally tend to infinity, which is not physical). Whether the model holds between these two extreme limits, however, remains to be determined using 3D DNS. 

\subsection{Numerical simulations of diffusive stratified shear instabilities}

Numerical simulations are therefore required to test the two parts of Zahn's model. The first three-dimensional DNS of stratified shear instabilities in the diffusive limit (more specifically, in the limit of low P\'eclet number) were presented by \citet{PratLignieres13} \citep[see also][]{PratLignieres14}, followed by \citet{Garaudal15}, \citet{GaraudKulen16} and \citet{Pratal2016}. In all cases, they apply codes that use a horizontally-periodic Cartesian domain, under the Boussinesq approximation. In nearly all cases \citep[except in parts of the work of][]{PratLignieres14}, chemical species are mixed as passive tracers and stratification is due to temperature alone, which is also the limit we consider here. The background temperature gradient is assumed to be constant (in the low P\'eclet number limit), and therefore so is the buoyancy frequency $N$. 

The works of Prat and his collaborators use two different model setups in which the imposed shear flow is linear everywhere in the domain, with a given constant shearing rate $S$. By doing so, they are testing Zahn's model as it was originally intended, namely as a {\it local} mixing model. To drive this linear shear flow, \citet{PratLignieres13}  and \citet{PratLignieres14} add a body force to the momentum equation that takes the form of a relaxation term driving the horizontal average of the streamwise component of the velocity field towards the linear target flow $Sz$, where $z$ is the vertical coordinate, on a certain timescale $\tau$. \citet{PratLignieres14} argued that the results are independent of the exact value of $\tau$ provided the latter is small enough. More recently, \citet{Pratal2016} revisited the homogeneously sheared model using the shearing sheet formalism instead. Crucially, they found that their results are consistent with those of \citet{PratLignieres13} and \citet{PratLignieres14}, hence validating the results of these studies, and showing that the manner in which the constant shear is forced does not influence the turbulent dynamics induced. 

\citet{Pratal2016} find that, in the low P\'eclet number limit, finite amplitude instabilities exist up to about $J {\rm Pr} \simeq 0.007$, but that initial perturbations decay for larger Richardson-Prandtl numbers. This constrains one of the unknown model constants, namely
\begin{equation}
(J{\rm Pr})_c \simeq 0.007\, ,
\end{equation}
a value that is quite close to the one predicted by \citet{Zahn1974}, namely $(J{\rm Pr})_c  = O(10^{-3})$. They also measure the turbulent mixing coefficient $D_{\rm turb}$ and use it to test the validity of the mixing prescription proposed in (\ref{eq:zahndt}). They find that $D_{\rm turb} J / \kappa_T$, which should be constant and equal to $C$, actually varies quite significantly with $J\Pr$, and somewhat with Reynolds number ${\rm Re}_L$. To be more specific, $D_{\rm turb} J / \kappa_T$ first gradually increases with $J\Pr$, reaches a maximum, then drops sharply to zero as $J{\rm Pr} \rightarrow (J\Pr)_c$ (see Figure \ref{fig:DtvsJPr}a in this paper for an equivalent figure made with our data). Meanwhile at low stratification $D_{\rm turb} J / \kappa_T$ increases with $\Re_L$, but becomes independent of it at larger stratification. 

While the sharp drop in $D_{\rm turb} J / \kappa_T$ as  $J{\rm Pr} \rightarrow (J\Pr)_c$ simply reflects the expected stabilization of the system against turbulence, \citet{Pratal2016} have more difficulty explaining the decrease of $D_{\rm turb} J / \kappa_T$ as $J\Pr$ decreases. Looking at the turbulent eddies, however, they find that their vertical size increases as $J\Pr$ decreases, and eventually approaches the domain size, at which point one should indeed expect Zahn's local model (\ref{eq:zahndt}) to fail. They argue this also explains the observed dependence of $D_{\rm turb}J / \kappa_T$ on the Reynolds number in the limit of low stratification. Indeed, since a larger Reynolds number corresponds to an intrinsically larger domain, it is possible to run simulations at lower values of $J\Pr$ without being constrained by the domain size. Extrapolating their results to much larger Reynolds numbers (which are more appropriate of stellar conditions), \citet{Pratal2016} then argue that $D_{\rm turb} J / \kappa_T$ should in reality tend to a constant in the limit of very low stratification. They conclude by proposing the following prescription as a refinement of Zahn's model that fits their data in the limit of large Reynolds number: 
\begin{equation}
\frac{D_{\rm turb} J}{\kappa_T} \simeq \alpha_0 + \alpha_1 J \Pr - \alpha_2 (J \Pr)^2\, ,
\label{eq:Pratmodel}
\end{equation}
where $\alpha_0 = 3.34 \times 10^{-2}$, $\alpha_1 = 18.8$ and $\alpha_2  = 2.86 \times 10^3$ \citep{Pratal2016}.  

In an attempt to move beyond homogeneous shear flows and investigate more complex shear profiles, Garaud and her collaborators \citep{Garaudal15,GaraudKulen16} chose to use a different model setup in which the shear is driven by a time-independent but sinusoidally varying body force ${\bf F}(z) = F_0 \sin(kz){\bf e}_x$, where $k$ is a vertical wavenumber and $F_0$ is the amplitude of the force. By contrast with the work of \citet{PratLignieres13}, \citet{PratLignieres14}, and \citet{Pratal2016}, the shear is not specified in their simulations but is instead free to evolve in time either viscously, or in response to the Reynolds stresses produced in the turbulent shear flow. As a result, once a quasi-stationary state has been achieved, the mean shear $S$ varies with $z$ and so does the gradient Richardson number $J(z)$.

Measuring $J(z)$ deep in the middle of the turbulent layer, \citet{GaraudKulen16} found that turbulent solutions exist up to $J {\rm Pr} \simeq 0.006$, but not for significantly higher values of $J\Pr$. Furthermore, they found that the mixing coefficient (\ref{eq:zahndt}) proposed by \citet{Zahn92} correctly models the transport of a passive scalar for a range of stratification, but fails either when it is too strong (due to the stabilization of the flow) or when it becomes too weak. In that case, they find that the size of the turbulent eddies becomes controlled by the lengthscale of the shear $k^{-1}$. All of these findings are entirely consistent with those of \citet{Pratal2016}. 

A new coherent picture of diffusive stratified shear instabilities thus emerges from the combination of all these results. The stability criterion of \citet{Zahn1974}, given in (\ref{eq:crit2}), was verified to hold, and the constant $(J \Pr)_c$ was measured in several independent ways to be $ (J \Pr)_c \simeq 0.007$. The mixing prescription of \citet{Zahn92}, given in (\ref{eq:zahndt}), also appears to hold for intermediate values of the stratification. It however fails in the limit of large stratification where the flow is stable, and in the limit of very low stratification where the turbulent eddy sizes become constrained either by the size of the domain (in the homogeneously sheared case) or by the shear lengthscale -- in other words, when the local assumption at the heart of Zahn's models fails. 

A few loose ends remain to be investigated, however. First, while the limit of large stratification is reasonably well understood, the limit of low stratification remains to be studied in more detail. Is it possible, for instance, to extend Zahn's mixing model \citep{Zahn92} to account for it as well? Second, both \citet{GaraudKulen16} and \citet{Pratal2016} attempted to test Zahn's mixing prescription more directly by measuring the dominant turbulent eddy scale $l_e$ and  comparing $S l_e^2$ with $D_{\rm turb}$ in the simulations. Both of these quantities are measured entirely independently (the first one being extracted from the power spectrum of the velocity fluctuations and the second one being extracted from the correlation between the velocity and passive tracer fields), and should be proportional to one another according to (\ref{eq:dturblz}) for Zahn's model to be correct. Curiously, \citet{GaraudKulen16} and \citet{Pratal2016} both found that $S l_e^2$ and $D_{\rm turb}$ are {\it not} proportional to one another, even in the case of intermediate stratification where (\ref{eq:zahndt}) roughly holds. This raises a strange conendrum \citep{GaraudKulen16}: how can a model hold if its fundamental assumption does not? 

To address these remaining questions, and look at the problem from yet a new angle, we study in this paper DNS of stratified plane Couette flows (i.e. stratified shear flows driven by plane-parallel no-slip plates moving in opposite directions) in the low P\'eclet number limit. In these numerical experiments, as we shall demonstrate, the shear flow between the plates is free to adjust itself (as in the body-forced case of Garaud and collaborators), but remains close to being linear in the bulk of the domain (as in the work of Prat and collaborators). This setup offers the advantage of being able to compare our results more easily with those of \citet{Pratal2016} than with the sinusoidally forced case. It is also motivated physically in the astrophysical context by the fact that many shear layers are forced by surface stresses rather than body forces. For instance, magnetic braking only affects the outer convective layers of a star, which then communicate the spin-down torque to the underlying radiative layer through interfacial stresses at the base of the convection zone. As another example, mass accretion onto a star, whether from a disk or a companion, usually deposits a thin layer of rapidly rotating fluid on the surface, which can also be viewed as imposing a surface stress to the underlying regions.

In what follows, we therefore present the stratified plane Couette flow setup in Section \ref{sec:model}, and analyze the simulations in Section \ref{sec:nums}. We then revisit Zahn's two-part model \citep{Zahn1974,Zahn92} in Section \ref{sec:analysis}, and propose a new physically-motivated mixing prescription for turbulent mixing by stratified shear instabilities that fits our numerical data in all limits. We conclude in Section \ref{sec:ccl} by summarizing our results and discussing avenues for future work.

\section{The model}
\label{sec:model}

The plane Couette flow is perhaps the simplest possible model setup for boundary-forced shear flows.  Two no-slip horizontal infinite parallel plates, separated by a distance $L$, move opposite one another other with velocities $\pm \frac{1}{2}\Delta U \be_x$, driving a shear flow between them. If this flow remains entirely laminar, then it  takes the form $\bar \bU_{\rm C} (z) = \frac{\Delta U}{L} z \be_x $, with a constant shearing rate $S_{\rm C} = \Delta U / L$, the subscript C standing for ``Couette". If we further assume that the top and bottom plates are held at constant temperatures $T_m + \Delta T/2$ and $T_m - \Delta T/2$, in the absence of temperature fluctuations, the background temperature gradient is constant and equal to $T_{0z} = \Delta T / L$. If in addition $T_{0z} > T^{\rm ad}_z$ where $T^{\rm ad}_z$ is the adiabatic temperature gradient, this system is stably stratified.  The dimensional equations and boundary conditions describing the evolution of the velocity field  $\bu$ and temperature fluctuations $T$ away from the background temperature profile $T_0(z) = T_m + T_{0z}z$, in the Boussinesq limit \citep{SpiegelVeronis1960}, are
\begin{eqnarray}
\frac{\partial \bu}{\partial t}+  \bu \cdot {\bf \nabla} \bu =-  \frac{1}{\rho_m} {\bf \nabla}  p + \alpha g T \be_z + \nu \nabla^2  \bu\, ,  \nonumber \\
{\bf \nabla} \cdot  \bu = 0 \, ,\nonumber \\
\frac{\partial  T}{\partial t}+  \bu \cdot {\bf \nabla}  T +  w (T_{0z}-T^{\rm ad}_z)  = \kappa_T \nabla^2   T\, , \label{eq:fullcouettedim}
\end{eqnarray}
and
\begin{eqnarray}
v = w =  T = 0 \mbox{ at } z = \pm L/2\, , \nonumber \\
u  = \pm \Delta U/2 \mbox{ at } z = \pm L/2\, ,
\end{eqnarray} 
together with periodic boundary conditions in the horizontal direction. In these equations, $\bu = (u,v,w)$ is the full 3D velocity field, $p$ is the pressure perturbation away from hydrostatic equilibrium, $\rho_m$ is the mean density of the fluid, $\alpha = - \rho_m^{-1}(\partial \rho/\partial T)_{p}$ is the coefficient of thermal expansion, and $g$ is gravity. The quantities $\alpha$, $g$, $\nu$, $\kappa_T$, $T_{0z}$ and $T^{\rm ad}_z$ are all assumed to be constant. 

We then non-dimensionalize these equations and boundary conditions assuming that the unit length is $L$, the unit time is $S_{\rm C}^{-1}$ and the unit temperature is $L (T_{0z}-T^{\rm ad}_z)$, to get 
\begin{eqnarray}
\frac{\partial \check \bu}{\partial t}+ \check \bu \cdot {\bf \nabla} \check\bu =-  {\bf \nabla} \check p + {\rm Ri}_{\rm C} \check T \be_z + \frac{1}{{\rm Re}_{\rm C}} \nabla^2\check \bu  \, ,\nonumber \\
{\bf \nabla} \cdot \check \bu = 0 \, ,\nonumber \\
\frac{\partial  \check T}{\partial t}+ \check \bu \cdot {\bf \nabla} \check T + \check w  = \frac{1}{ {\rm Pe}_{\rm C}} \nabla^2\check  T\, , \label{eq:fullcouette}
\end{eqnarray}
and
\begin{eqnarray}
\check v = \check w = \check T = 0 \mbox{ at } z = \pm 1/2\, , \nonumber \\
\check u  = \pm 1/2 \mbox{ at } z = \pm 1/2\, ,
\end{eqnarray} 
where pressure, the spatio-temporal coordinates, and the differential operators have also been implicitly non-dimensionalized, and where
\begin{equation}
\Re_{\rm C} = \frac{S_{\rm C} L^2}{\nu}, \, \, \Pe_{\rm C} = \frac{S_{\rm C} L^2}{\kappa_T}, \, \, \mbox{ and } {\rm Ri}_{\rm C} = \frac{ \alpha g (T_{0z}-T^{\rm ad}_z)}{S_{\rm C}^2} = \frac{N^2}{S_{\rm C}^2}\, ,
\end{equation}
are the Reynolds, P\'eclet and bulk Richardson numbers based on the laminar shear $S_{\rm C}$. 

The asymptotic low P\'eclet number approximation \citep{Lignieres1999} of the equations listed in (\ref{eq:fullcouette}), which we shall use in this paper, is given by  
\begin{eqnarray}
\frac{\partial \check \bu}{\partial t}+ \check \bu \cdot {\bf \nabla} \check\bu =-  {\bf \nabla} \check p + \Ri_{\rm C} \check T \be_z + \frac{1}{{\rm Re}_{\rm C}} \nabla^2\check \bu\, ,  \nonumber \\
{\bf \nabla} \cdot \check \bu = 0\, , \nonumber \\
\check w  = \frac{1}{ {\rm Pe}_{\rm C}} \nabla^2\check  T\, .  \label{eq:LPNcouette}
\end{eqnarray}
These equations are referred to as the LPN equations hereafter. Finally, we also follow the evolution of a nondimensional passive scalar field $\check c$ (representing for example the concentration of a given chemical species), according to
\begin{equation}
\frac{\partial  \check c}{\partial t}+ \check \bu \cdot {\bf \nabla} \check c   = \frac{\tau_c}{ {\rm Pe}_{\rm C}} \nabla^2\check  c\, , \label{eq:compo}
\end{equation}
where $\tau_c = \kappa_c / \kappa_T$ is the ratio of the microscopic compositional diffusivity to the thermal diffusivity, and where $\check c$ satisfies the following Dirichlet conditions: 
\begin{equation}
\check c  =  \pm 1/2 \mbox{ at } z = \pm 1/2\, .
\end{equation}
%XXX Note : these are not the actual BCs but they are completely equivalent and more symmetric XXX 
In all that follows, we take $\tau_c = 10^{-3}$, which is much larger than realistic values of the diffusivity ratio in stellar interiors (where it is closer to $10^{-9} - 10^{-7}$), but the smallest value we can realistically achieve in fully resolved DNS. With this choice, the chemical diffusivity is equal to the kinematic viscosity.

 Note that the parameters $\Re_{\rm C}$, $\Pe_{\rm C}$ and $\Ri_{\rm C}$ are different from $\Pe_L$, $\Re_L$ and $J$ defined and used by \citet{PratLignieres13}, \citet{PratLignieres14} and \citet{Pratal2016}, since the former are based on the laminar shearing rate $S_{\rm C}$ while the latter are based on the actual shearing rate $S$. As such, one cannot directly compare the results of our simulations at fixed $\Re_{\rm C}$ and $\Ri_{\rm C} \Pe_{\rm C}$ with those of their simulations at the same values of $\Re_L$ and $J \Pe_L$ respectively. However, once the actual value of the non-dimensional shear $\check S = S/S_{\rm C}$ in the bulk of the fluid between the plates is measured in our plane Couette simulations, we can calculate what the equivalent values of $\Pe_L$, $\Re_L$ and $J$ would be using
\begin{equation}
\Pe_L = \frac{S L^2}{\kappa_T } = \check S \Pe_{\rm C} \mbox{    ,   } \Re_L = \frac{S L^2}{\nu }  = \check S \Re_{\rm C} \mbox{    and  }  J = \frac{N^2}{S^2} = \frac{\Ri_{\rm C}}{\check S^2}\, ,
\label{eq:Jdef1}
\end{equation}
and use these to compare our results to those of Prat and collaborators. 

\section{Numerical simulations}
\label{sec:nums}

We solve the set of equations given in (\ref{eq:LPNcouette}) in a cubic domain of non-dimensional side-length $L = 1$ using a code developed by \citet{Verhoeven2014143} \citep[see also][]{Verhoeven2015}, modified to include the moving boundaries and to solve the LPN equations (\ref{eq:LPNcouette}) together with (\ref{eq:compo}). The code is doubly-periodic and therefore spectral in the horizontal directions, and uses fourth-order finite differences in the vertical direction. We verified that the LPN equations are a good representation of the dynamics of the full equations (\ref{eq:fullcouette}) in the limit of low $\Pe_{\rm C}$, as in \citet{GaraudKulen16}. Table \ref{table1} summarizes our results.

\begin{table}[]
	\caption{Summary of the main results for all the runs. The first column reports $\Re_{\rm C}$, the second column reports $\Ri_{\rm C} \Pe_{\rm C}$. The third column is the number of actual or equivalent mesh points in each direction. The fourth column is the amplitude of the mean nondimensional shear in the middle of the domain, the fifth is the amplitude of the mean nondimensional concentration gradient in the middle of the domain, the sixth is the eddy lengthscale $l_e$, the seventh is the time-averaged and volume-averaged turbulent compositional flux and the eight is the time-averaged and volume-averaged Reynolds stress.}
	\label{table1}
	\centering{
	\vspace{0.3cm}
	%\resizebox{\linewidth}{!}{\begin{tabular}{cccccccc}
\begin{tabular}{cccccccc}
		\tableline
		$\Re_{\rm C}/10^4$ & $\Ri_{\rm C} \Pe_{\rm C}$  & $ N_x, N_y, N_z$  & $\check S$        &       $\check G$       &      $l_e$   &    $ \langle \check w \check c \rangle_t \times 10^4 $ &         $ \langle \check w \check u \rangle_t \times 10^4$                 \\
			\tableline
	4 & $2.5 \times 10^{-4}$ & $96^2 \times 128$ & $0.235\pm0.003$& $0.22\pm0.11$ & 0.178 & $7.12\pm1.4$  & $6.12\pm 2.8$   \\
	4 & $2.5 \times 10^{-3}$ & $96^2 \times 128$ & $0.221\pm0.003$& $0.22\pm0.11$ & 0.162 & $7.12\pm1.4$  & $6.12\pm 2.8$   \\
	4 & $2.5 \times 10^{-2}$ & $96^2 \times 128$ & $0.246\pm0.003$& $0.23\pm0.11$ & 0.178 & $7.08\pm1.4$  & $6.12\pm 2.8$   \\
	4 & $2.5 \times 10^{-1}$ & $96^2 \times 128$ & $0.25\pm0.003$& $0.23\pm0.11$ & 0.162 & $5.88\pm1.2$ & $6.06\pm 2.6$ \\
	4 & $2.5 $ & $96^2 \times 128$ & $0.345\pm0.003$ & $0.29\pm0.11$ & 0.116 & $5.22\pm0.8$ & $5.18\pm 1.7$  \\
	4 & $8.25 $ & $96^2 \times 128$ & $0.461\pm0.002$ & $0.38\pm0.11$ & 0.093 & $4.32\pm0.50$ & $4.18\pm 1.1$ \\
	4 & $25 $ & $96^2 \times 128$ & $0.54\pm0.001$ & $0.48\pm0.11$ & 0.077 & $3.37\pm0.27$ & $3.25\pm 0.6$ \\
	4 & $82.5 $ & $96^2 \times 128$ & $0.68\pm0.001$ & $0.67\pm0.12$ & 0.062 & $1.99\pm0.14$ & $2.01\pm 0.3$ \\
	4 & $165 $ & $96^2 \times 128$ & $0.79\pm0.001$ & $0.82\pm0.13$ & 0.054 & $1.03\pm0.10$ & $1.10\pm 0.3$ \\
\\
	6 & $1.66 \times 10^{-2}$ & $144^2 \times 192$ & $0.224\pm0.004$& $0.21\pm0.13$ & 0.27 & $5.50\pm1.0$ & $5.58\pm 2.3$   \\
	6 & $1.66 \times 10^{-1}$ & $144^2 \times 192$ & $0.224\pm0.006$& $0.24\pm0.15$ & 0.15 & $5.38\pm0.9$  & $5.55\pm 2.1$   \\
	6 & $1.66 $ & $144^2 \times 192$ & $0.313\pm0.004$& $0.28\pm0.15$ & 0.10 & $4.91\pm0.8$  & $4.89\pm 1.2$   \\
	6 & $16.6 $ & $144^2 \times 192$ & $0.505\pm0.002$& $0.44\pm0.15$ & 0.078 & $3.41\pm0.3$  & $3.2\pm 0.6$   \\
	6 & $50 $ & $144^2 \times 192$ & $0.602\pm0.001$& $0.55\pm0.15$ & 0.067 & $2.48\pm0.1$  & $2.3\pm 0.3$   \\
	6 & $166 $ & $144^2 \times 192$ & $0.746\pm0.001$& $0.72\pm0.15$ & 0.047 & $1.25\pm0.07$  & $1.3\pm 0.2$   \\
\\
	9 & $1.11 \times 10^{-3}$ & $192^2 \times 256$ & $0.225\pm0.004$& $0.21\pm0.11$ & 0.22 & $5.28\pm1.0$  & $5.32\pm 1.9$   \\
	9 & $1.11 \times 10^{-2}$ & $192^2 \times 256$ & $0.191\pm0.003$& $0.20\pm0.11$ & 0.22 & $5.39\pm0.9$  & $5.36\pm 2.4$   \\
	9 & $1.11 \times 10^{-1}$ & $192^2 \times 256$ & $0.214\pm0.002$& $0.21\pm0.08$ & 0.15 & $5.21\pm1,0$  & $5.21\pm 2.1$   \\
	9 & $1.11 $ & $192^2 \times 256$ & $0.297\pm0.002$& $0.27\pm0.08$ & 0.12 & $4.77\pm0.8$  & $4.71\pm 1.7$   \\
	9 & $11.1 $ & $192^2 \times 256$ & $0.480\pm0.001$& $0.40\pm0.09$ & 0.085 & $3.38\pm0.3$  & $3.16\pm 0.6$   \\
	9 & $111 $ & $192^2 \times 256$ & $0.686\pm0.0003$& $0.64\pm0.09$ & 0.047 & $1.67\pm0.07$  & $1.56\pm 0.2$   \\
	9 & $333 $ & $192^2 \times 256$ & $0.812\pm0.0002$& $0.80\pm0.09$ & 0.035 & $0.84\pm0.03$  & $0.74\pm 0.07$   \\
	\\
	12 & $8.33 \times 10^{-3}$ & $192^2 \times 256$ & $0.204\pm0.004$& $0.20\pm0.1$ & 0.19 & $4.98\pm0.9$  & $4.93\pm 1.9$   \\
	12 & $8.33\times 10^{-2}$ & $192^2 \times 256$& $0.202\pm0.003$& $0.20\pm0.08$ & 0.20 & $5.10\pm1.1$  & $5.10\pm 2.3$   \\
	12 & $8.33 \times 10^{-1}$ & $192^2 \times 256$ & $0.241\pm0.003$& $0.23\pm0.09$ & 0.15 & $4.81\pm0.9$  & $4.75\pm 1.8$   \\
	12 & $8.33$ & $192^2 \times 256$ & $0.464\pm0.001$& $0.38\pm0.08$ & 0.097 & $3.50\pm0.3$  & $3.21\pm 0.7$   \\
	12 & $83.3$ &$192^2 \times 256$ & $0.655\pm0.0003$& $0.59\pm0.09$ & 0.05 & $1.83\pm0.07$  & $1.65\pm 0.2$   \\
	12 & $250$ & $192^2 \times 256$ & $0.762\pm0.001$& $0.74\pm0.09$ & 0.039 & $1.05\pm0.02$  & $1.00\pm 0.09$   \\
	12 & $500$ & $192^2 \times 256$& $0.850\pm0.0001$& $0.84\pm0.06$ & 0.031 & $0.51\pm0.01$ & $0.52 \pm 0.05$   \\
\\

\end{tabular}
}
\end{table}

 The code being originally written for studies of Rayleigh-B\'enard convection, the numerical input parameters are the Rayleigh number $\Ra = \Ri_{\rm C} \Pe_{\rm C} \Re_{\rm C}$, $\Pr = \Pe_{\rm C}/ \Re_{\rm C}$ and $\Re_{\rm C}$. As a result, our parameter sweeps are at constant $\Ra$ and constant $\Re_{\rm C}$ rather than at constant $\Ri_{\rm C} \Pe_{\rm C}$ and constant $\Re_{\rm C}$. Nevertheless, we can still use the data to make useful comparisons between runs. Finally, in all that follows, we use the notations 
\begin{eqnarray}
\overline{q}(z,t) = \frac{1}{L^2} \iint q(x,y,z,t)  dx dy\, , \\
\langle q \rangle(t) =  \frac{1}{L^3} \iiint q(x,y,z,t)  dx dy dz\, , \\
(q)_t(x,y,z) = \frac{1}{t_2 - t_1} \int_{t_1}^{t_2} q(x,y,z,t) dt \, , 
\end{eqnarray}
to imply horizontal-, volume- and time-averages respectively. In the expression for the time-average, the times $t_1$ and $t_2$ are selected to bracket the longest possible time-interval available in the simulations during which the shear flow has achieved a statistically-stationary state.  

\subsection{Sample results}

\begin{figure}[h]
  \centerline{\includegraphics[width=0.45\textwidth]{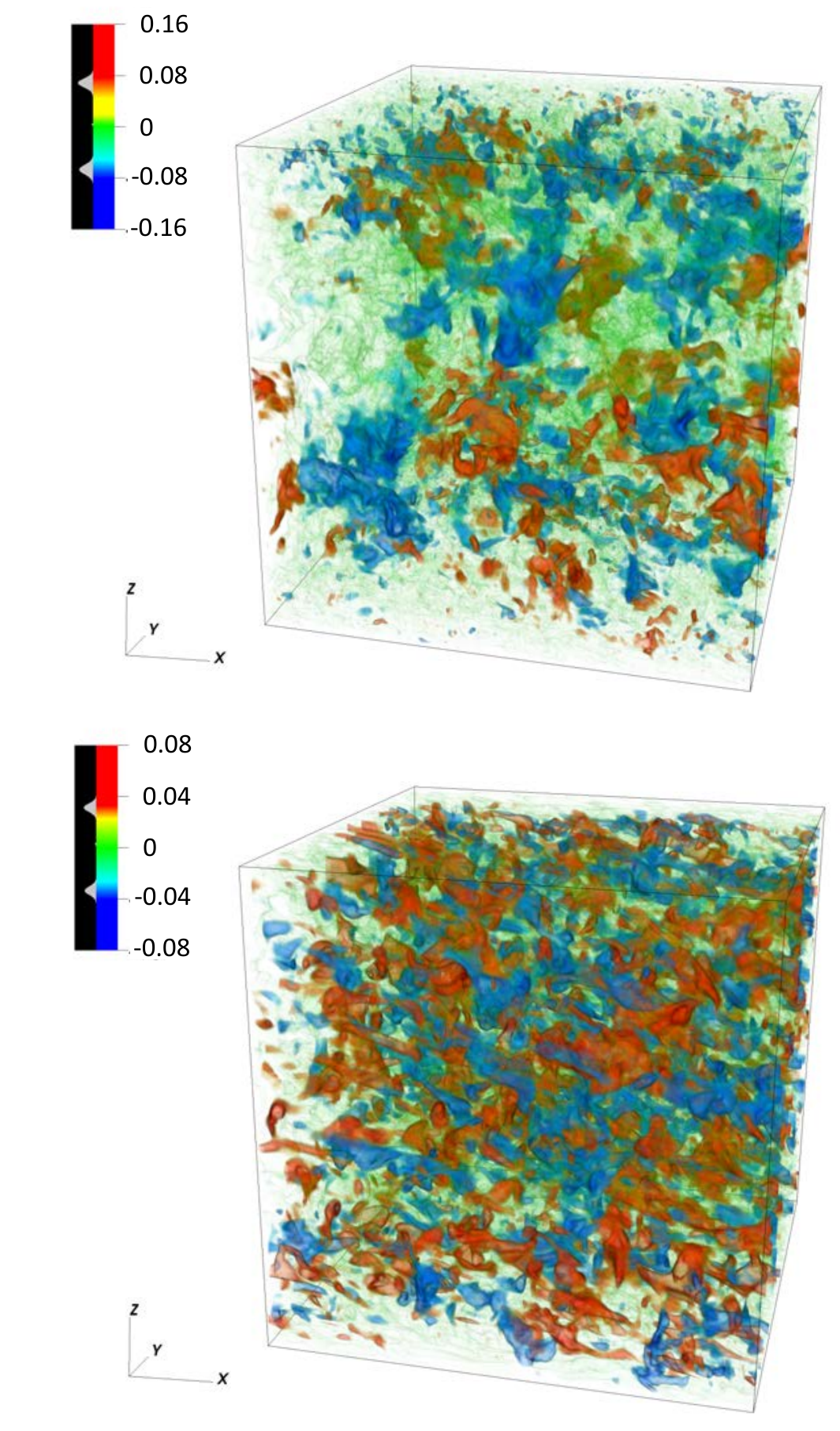}}
  \caption{Volume rendered snapshots of the vertical velocity field $\check w$ in two simulations at $\Re_{\rm C} = 1.2 \times 10^5$. Top: $\Ri_{\rm C}\Pe_{\rm C} =$ 0.0833. Bottom:  $\Ri_{\rm C}\Pe_{\rm C} =$ 83.3.}
\label{fig:prettypics}
\end{figure}

Figure \ref{fig:prettypics} shows volume-rendered snapshots of the vertical velocity field in two simulations at $\Re_{\rm C} = 1.2 \times 10^5$, and $\Ri_{\rm C}\Pe_{\rm C} =$  0.0833  and $83.3$ respectively. Note how the typical vertical coherence of the eddies is much larger in the more weakly stratified case (at lower $\Ri_{\rm C}\Pe_{\rm C}$) than in the more strongly stratified one, even though it retains significant fine-scale structure owing to the large Reynolds number of the simulations. 

\begin{figure}[h]
  \centerline{\includegraphics[width=\textwidth]{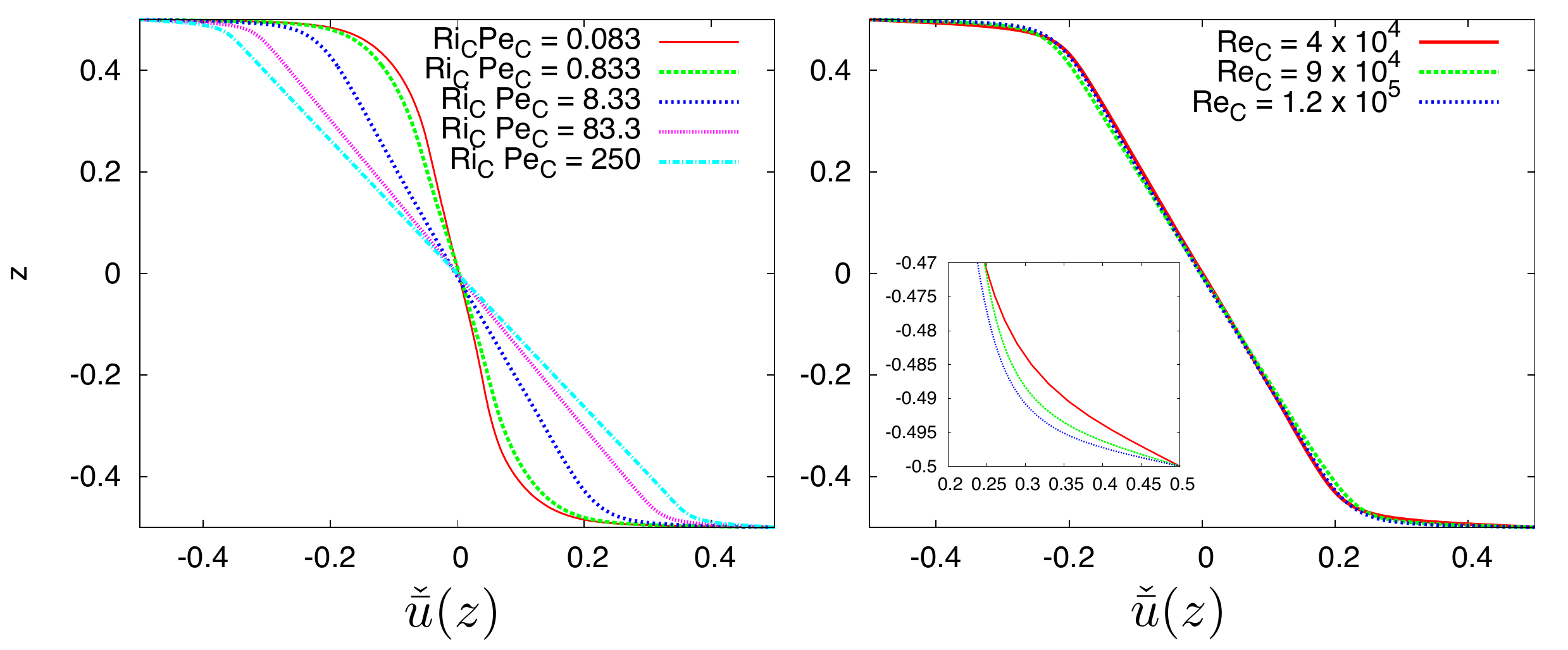}}
  \caption{Left: horizontally-averaged and time-averaged mean flow $(\check{\bar u})_t$ as a function of $z$, for various simulations at $\Re_{\rm C} = 1.2 \times 10^5$.  Right: Same quantity, but this time varying $\Re_{\rm C}$ and holding $\Ri_{\rm C} \Pe_{\rm C}$ roughly constant. The runs shown have $\Ri_{\rm C} \Pe_{\rm C} = 8.25$, $11.1$ and $8.33$ respectively as $\Re_{\rm C}$ increases. }
\label{fig:meanflow1}
\end{figure}

\begin{figure}[h]
  \centerline{\includegraphics[width=\textwidth]{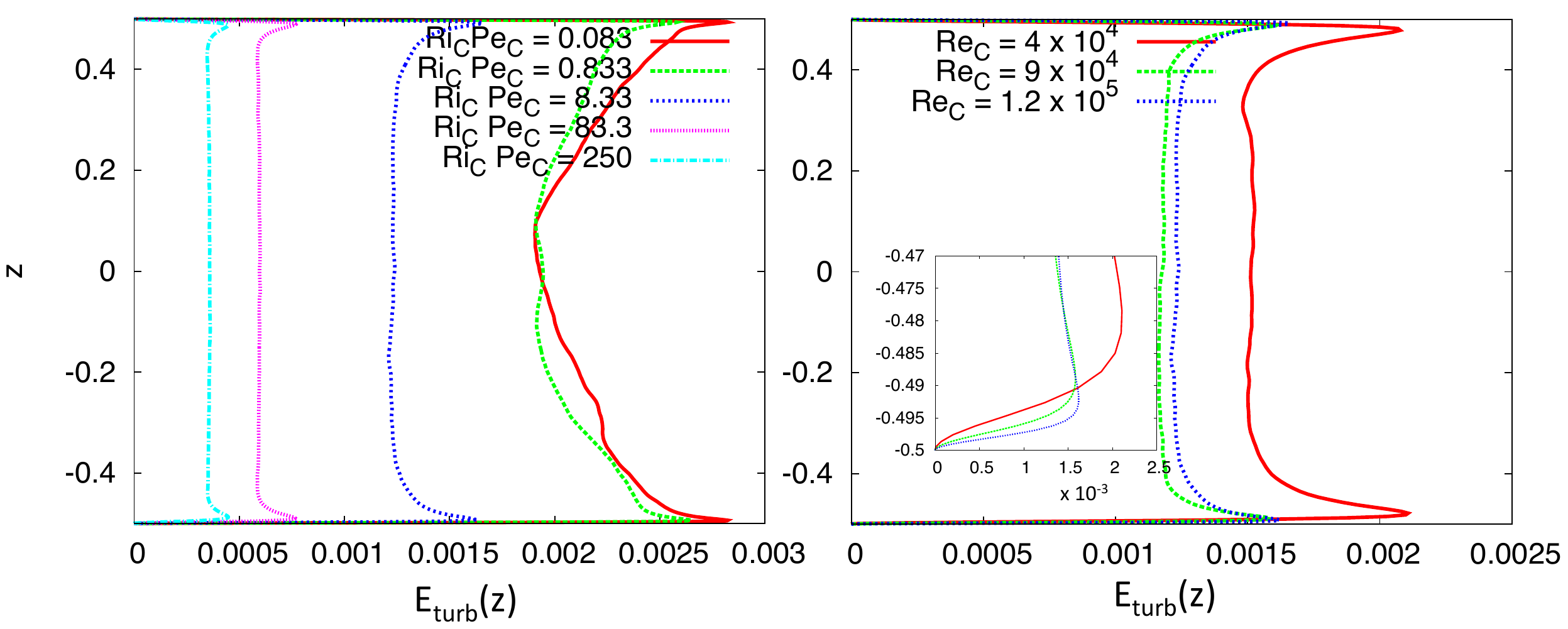}}
  \caption{Left: Horizontally-averaged and time-averaged turbulent kinetic energy $E_{\rm turb}$ as a function of $z$, for the same simulations as in Figure \ref{fig:meanflow1}. Right: Same quantity, but this time varying $\Re_{\rm C}$ and holding $\Ri_{\rm C} \Pe_{\rm C}$ roughly constant. The runs shown have $\Ri_{\rm C} \Pe_{\rm C} = 8.25$, $11.1$ and $8.33$ respectively as $\Re_{\rm C}$ increases. }
\label{fig:kin1}
\end{figure}

In order to analyze the system more quantitatively, Figure \ref{fig:meanflow1} shows the horizontally-averaged and time-averaged mean flow $(\check{\bar u})_t$ for two sets of simulations: varying $\Ri_{\rm C} \Pe_{\rm C}$ at constant $\Re_{\rm C}$ (left) and varying $\Re_{\rm C}$ at roughly constant $\Ri_{\rm C} \Pe_{\rm C}$ (right). Similarly, Figure \ref{fig:kin1} shows the horizontally-averaged and time-averaged turbulent kinetic energy, 
\begin{equation}
E_{\rm turb}(z) = \frac{1}{2} \left( \overline{\check u^2} - \check{\bar u}^2 + \overline{\check v^2} + \overline{\check w^2} \right)_t 
\end{equation}
as a function of $z$ in the same sets of simulations. Both figures show that the computational domain is divided into a bulk region where the time-averaged shear and turbulent kinetic energy are roughly constant, and two thin boundary layers on either side where the mean flow velocity is forced to match the wall velocity, and where the turbulent kinetic energy first grows somewhat then drops to zero to match the boundary conditions. The tendency for more strongly stratified turbulent systems to have large regions of constant shear is interestingly reminiscent of the numerical results of \citet{GaraudKulen16}, and will be discussed in more detail in a forthcoming publication. 

As the stratification (measured via $\Ri_{\rm C} \Pe_{\rm C}$) increases while holding $\Re_{\rm C}$ constant, the shear in the bulk of the domain also increases while the turbulent kinetic energy decreases.  Meanwhile for fixed $\Ri_{\rm C} \Pe_{\rm C} \simeq 8.5$, increasing the Reynolds number decreases the size of the viscous boundary layer but does not seem to affect the strength of the bulk shear. %Increasing $\Re_{\rm C}$ does seem to affect the turbulent kinetic energy in the bulk of the fluid, although there is some evidence that some convergence to a universal profile independent of viscosity may be achieved for large enough $\Re_{\rm C}$ (except for the thin boundary layers as in the case of the mean shear).

\begin{figure}[h]
  \centerline{\includegraphics[width=0.5\textwidth]{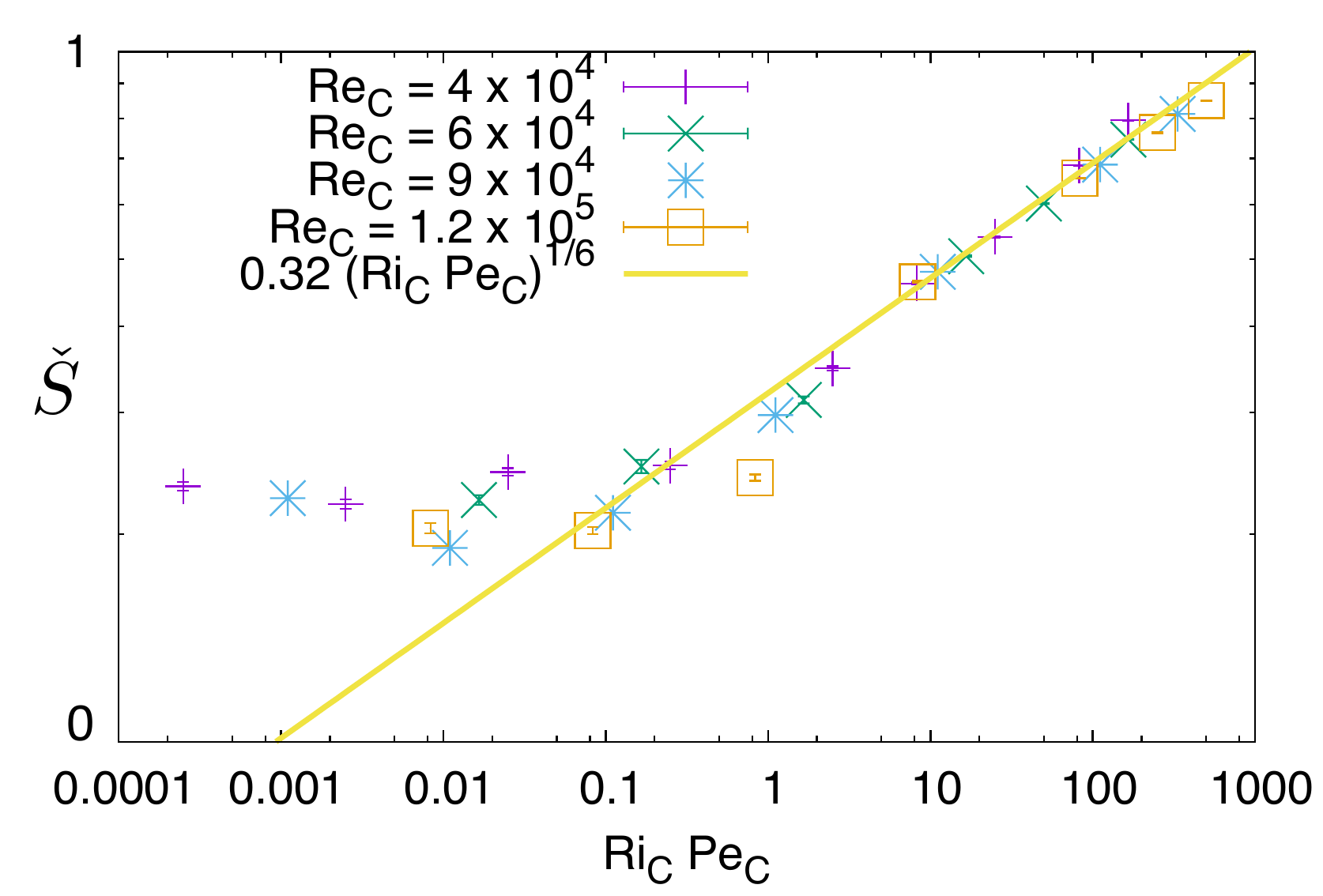}}
  \caption{Variation of the non-dimensional bulk shearing rate $\check S$ with $\Ri_{\rm C}\Pe_{\rm C}$. Various symbols correspond to various Reynolds numbers, as seen in the legend. A laminar flow would have $\check S = 1$. For low $\Ri_{\rm C}\Pe_{\rm C}$, $\check S$ shows some variation but is overall fairly constant and roughly equal to 0.2. The thick yellow line shows the high $\Ri_{\rm C}\Pe_{\rm C}$ scaling law. }
\label{fig:Jcompare}
\end{figure}

Figure \ref{fig:Jcompare} shows the amplitude of the nondimensional shear in the middle of the domain $\check S$, as a function of the input Richardson-P\'eclet number $\Ri_{\rm C} \Pe_{\rm C}$ and of $\Re_{\rm C}$. To compute $\check S$ we take the time- and horizontally-averaged profiles $(\check{\bar u})_t(z)$ shown in Figure \ref{fig:meanflow1}, and fit each of them to the linear function $f(z) = - \check Sz$ for $z \in [-0.1,0.1]$. The shear values thus extracted, together with its rms fluctuations, are also reported in Table 1. Note that, with this definition, $\check S$ is equal to {\it minus} the gradient of $(\check{\bar u})_t(z)$. We see that there are actually two distinct regimes, for low and high stratification respectively. For very low stratification (as measured by $\Ri_{\rm C} \Pe_{\rm C} \ll 1$), $\check S$ is more-or-less independent of both $\Re_{\rm C}$ and $\Ri_{\rm C} \Pe_{\rm C}$.  As $\Ri_{\rm C} \Pe_{\rm C}$ increases beyond one, the effects of stratification clearly become important and a new scaling law emerges with $\check S\simeq 0.32 (\Ri_{\rm C} \Pe_{\rm C})^{1/6}$. The origin of this law remains to be determined, but likely depends on the structure of the boundary layers near the walls, and is therefore probably model specific (i.e., in a different model that uses a different type of forcing, the relationship between the midpoint shear and the input Richardson-P\'eclet number is likely to be different). More importantly, however, we see that it is independent of the Reynolds number, which corroborates our observations of the bulk shear $\check S$ in Figure \ref{fig:meanflow1}. %As $\Ri_{\rm C} \Pe_{\rm C}$ continues to increase, $\check S \rightarrow 1$, which corresponds to the laminar shear solution. The turbulent solutions disappear altogether at this point. 

%The fact that this is precisely $1/{\rm Pr}$ is probably not a coincidence. In that case we would have 
%\begin{equation}
%J_C \Pe_C \simeq \frac{\Re_C}{\Pe_C}
%\end{equation}
%would be the critical number for instability. This is notably different from the finding of Prat et al. who find that the system becomes stable when $J \Pr$ approaches a critical value (similarly for Garaud et al.). This could suggest that there is something quite difference between body-forced and boundary-forced systems. That's not entirely surprising actually. For the boundary force ones there is a maximum value of the shear that the system can possibly achieve, while in the body-forced case the shear can in principle grow to any possible value. 

\subsection{Mixing}
\label{sec:mixing}

One of the goals of this investigation, as discussed in Section \ref{sec:intro}, is to quantify mixing by shear instabilities, both of a passive tracer and of momentum. A simple model for mixing consists in assuming that the turbulent fluxes are proportional to the local gradient of the mixed quantity, the coefficient of proportionality between the two being called turbulent diffusivity ($D_{\rm turb}$) or turbulent viscosity ($\nu_{\rm turb}$). In order to obtain good statistics, we take a time average and volume average of the turbulent fluxes, and let
\begin{eqnarray}
\check{D}_{\rm turb} =   \frac{\langle \check w \check c \rangle_t }{ \check G }\, , \label{eq:dturb} \\
\check{\nu}_{\rm turb} =  \frac{\langle \check u \check w\rangle_t }{ \check S} \, ,  \label{eq:nuturb}
\end{eqnarray}
where $\check{D}_{\rm turb}  = D_{\rm turb}/ \Delta U L$ is the non-dimensional turbulent diffusivity (and similarly for $\check{\nu}_{\rm turb}$), and where $\check G $ is minus the gradient of composition in the middle of the domain, measured exactly as $\check S$ (see above) but using the time- and horizontally-averaged compositional profile $(\check{\bar c})_t(z)$ instead of $(\check{\bar u})_t(z)$. Note that the absence of the usual negative sign in these definitions comes from the fact that we have defined $\check G$ and $\check S$ to be equal to minus the gradients of $(\check{\bar c})_t(z)$ and $(\check{\bar u})_t(z)$ respectively.

\begin{figure}[h!]
  \centerline{\includegraphics[width=0.5\textwidth]{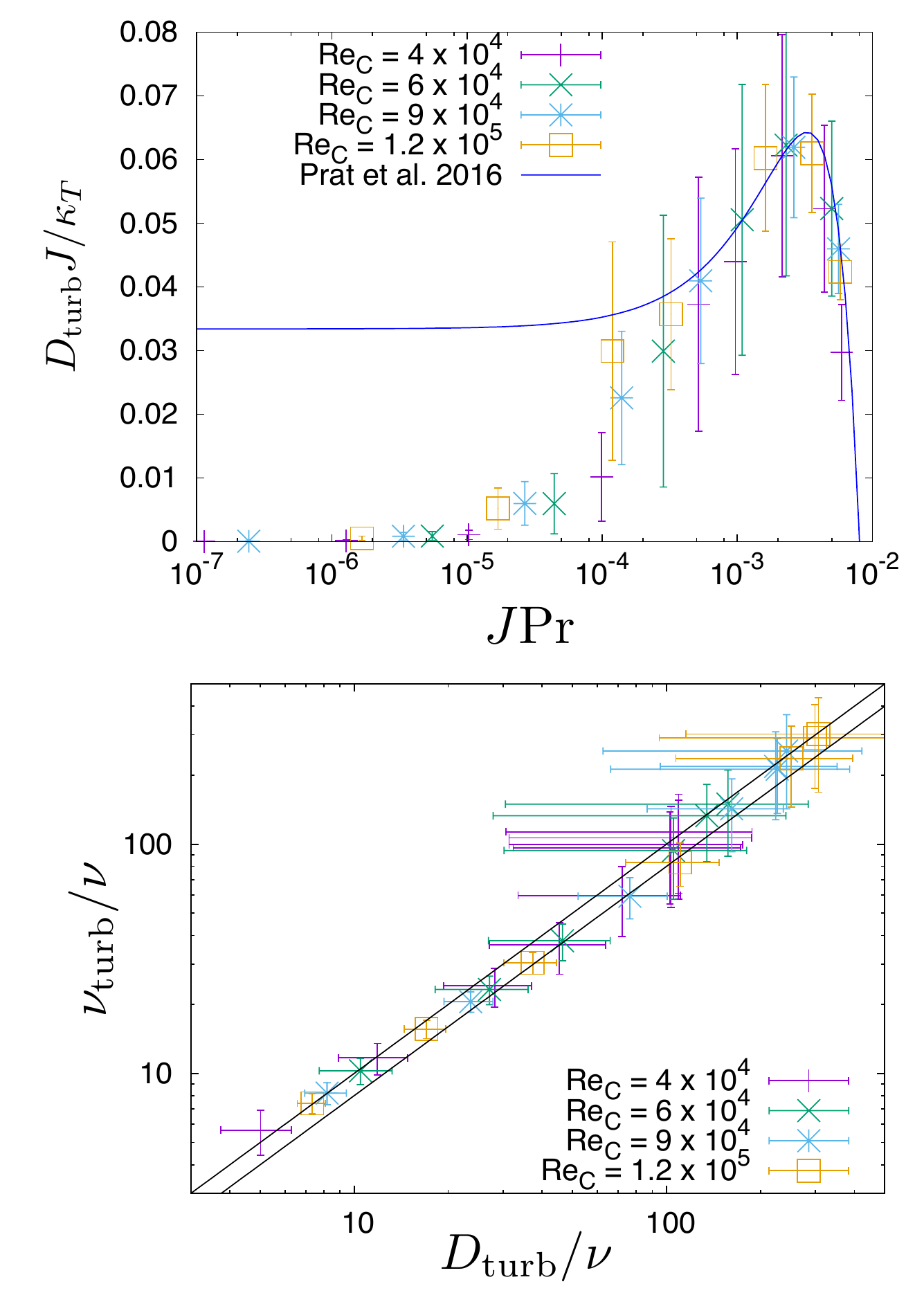}}
  \caption{These figures reproduce Figures 3 and 8 of \citet{Pratal2016}. Top: Variation of  $D_{\rm turb} J / \kappa_T = \check{D}_{\rm turb} J {\rm Pe}_{\rm_C}$ against $J \Pr$, for various Reynolds numbers. Also shown is the parametric model of \citet{Pratal2016} (see Equation \ref{eq:Pratmodel}), which fits the data well for the more strongly stratified simulations. Bottom: Variation of  $\nu_{\rm turb}/\nu = \check{\nu}_{\rm turb} \Re_{\rm_C}$ with $D_{\rm turb}/\nu = \check{D}_{\rm turb} \Re_{\rm_C}$. Note that this is shown on a log-log plot. Also shown are the relationships $\nu_{\rm turb} = 0.8 D_{\rm turb}$ and $\nu_{\rm turb} = D_{\rm turb}$. The former fits the data at intermediate values of the stratification, while the latter fits the data well for the more weakly and more strongly stratified simulations. Note that the fairly large errorbars come from relatively large rms fluctuations in $\check G$.}
\label{fig:DtvsJPr}
\end{figure}

Figures \ref{fig:DtvsJPr}a and \ref{fig:DtvsJPr}b use the calculated values of $\check{D}_{\rm turb}$ and $\check{\nu}_{\rm turb}$ to reproduce Figures 3 and 8 of \citet{Pratal2016}, albeit with the results of our own simulations. Figure \ref{fig:DtvsJPr}a shows the non-dimensional quantity $D_{\rm turb} J / \kappa_T = \check{D}_{\rm turb} J {\rm Pe_C}$ as a function of $J \Pr$, where $J$ is calculated from $\check S$ as given in (\ref{eq:Jdef1}). We also show, as a solid line, the parametric model of equation (\ref{eq:Pratmodel}), which was proposed by \citet{Pratal2016} to fit their data. 
%\citet{Pratal2016} noted that this quantity should be constant for Zahn's mixing model \citep{Zahn92} to hold, and should in particular be independent of the Reynolds number. They ran a suite of simulations varying $J \Pr$ between $10^{-4}$ and $10^{-2}$ -- a range that is somewhat more limited than ours -- and found that, instead, $D_{\rm turb} J / \kappa_T$ varies with $J \Pr$. They fitted their data to a parametric function, proposing that 
%independently of $\Re_C$.  \citet{Pratal2016} noted that their data deviates from their proposed universal function for low $J \Pr$, and that the deviations are more pronounced for lower Reynolds number. They argued that this is to be expected when the sizes of the turbulent eddies in the simulations become comparable with the domain size, at which point the model is artificially invalidated. 
We see that, for larger values of $J \Pr$, our data falls on their parametric curve. In particular, we find as they do that $D_{\rm turb} J / \kappa_T$ becomes independent of the Reynolds number for large $J\Pr$ and rapidly drops to zero as $J\Pr \rightarrow (J\Pr)_c \simeq 0.007$. This confirms the findings of \citet{Pratal2016}  that diffusive stratified shear instabilities can only occur as long as $J\Pr < (J\Pr)_c \simeq 0.007$. Note that this result, together with equation (\ref{eq:crit2}), provides an important experimental constraint on the constants $(J \Pe)_c$ and $\Re_c$. 

We also find, as they do, that the data deviates significantly from their parametric model in the limit of low $J \Pr$, and that the amplitude of the deviations increases with decreasing Reynolds number. In fact, we have ran a few cases with much lower values of $J \Pr$ than they did (their lowest value of $J \Pr$ was around $10^{-4}$, while ours is just below $10^{-6}$), and find that this trend continues as $J \Pr$ decreases. This confirms that their parametric model (and Zahn's model) fares poorly in the limit of low stratification, a result we expected from the discussion of Section \ref{sec:intromix}. We revisit the model with the goal of improving it in this limit in Section \ref{sec:analysis}. 

Figure \ref{fig:DtvsJPr}b shows $\nu_{\rm turb}/\nu = \check{\nu}_{\rm turb} \Re_{\rm_C}$ against $D_{\rm turb}/\nu = \check{D}_{\rm turb} \Re_{\rm_C}$, and illustrates that $\nu_{\rm turb}$ is indeed proportional to $D_{\rm turb}$, with a constant of proportionality that varies between 0.8 and 1. For comparison, \citet{Pratal2016} found that $\nu_{\rm turb} \simeq 0.8D_{\rm turb}$ for all values of the stratification, a result that is a little bit at odds with ours. This discrepancy may be due to the difference in the boundary conditions between the two codes: while they use a shearing-sheet without solid boundaries, our plane Couette flow model has solid boundaries. This introduces boundary layers near the walls (both in velocity and in the passive tracer field), that could act as a bottleneck to the overall transport rate and thus mildly affect $D_{\rm turb}$ and $\nu_{\rm turb}$. 

Aside from the slight discrepancy just mentioned, Figures \ref{fig:DtvsJPr}a and \ref{fig:DtvsJPr}b, and their resemblance with Figures 3 and 8 of \citet{Pratal2016}, confirm that the properties of the turbulent mixing coefficients $D_{\rm turb}$ and $\nu_{\rm turb}$ depend more on the properties of the local shear in the more strongly stratified limit, than on the manner in which the shear is forced (i.e. body forcing vs. forcing by surface stresses), the same conclusion reached by \citet{Pratal2016} when comparing two different methods of body-forcing (shearing sheet vs. explicit forcing). This is reassuring, and suggests that, at least in these cases, a local model for turbulent mixing by shear instabilities that only depends on the local gradient Richardson number $J$, together with the local microscopic diffusion coefficients $\nu$ and $\kappa_T$, is appropriate. 

\section{Analysis and discussion of the results}
\label{sec:analysis}

We now investigate our data in comparison with Zahn's model \citep{Zahn1974,Zahn92} in more depth, and attempt to resolve the outstanding issues described in Section \ref{sec:intro}. 

\subsection{Comparison of the various model lengthscales}
\label{sec:lengthscales}

As discussed in Sections \ref{sec:introstab} and \ref{sec:intromix}, Zahn's theory relies on the idea that turbulent eddies in diffusive stratified shear flows are just marginally unstable according to the criterion (\ref{eq:crit2}). This defines the Zahn scale $l_{\rm Z}$, to be (\ref{eq:zahnscale}):
\begin{equation}
l_{\rm Z} =  \left( \frac{(J \Pe)_c }{J \Pe_L }\right)^{1/2} L \, ,
\label{eq:zahnscale2}
\end{equation}
which can then be used in (\ref{eq:dturblz}) to estimate $D_{\rm turb}$ (and similarly $\nu_{\rm turb}$ since they are dimensionally equivalent). As implied by \citet{Zahn1974}, however,  for turbulence to exist $l_{\rm Z}$ has to be larger than the scale below which viscosity becomes important, which we call the viscous scale $l_\nu$. The value of $l_\nu$ can be derived from the condition that the eddy-scale Reynolds number is just equal to $\Re_c$ (see Section \ref{sec:introstab}):
\begin{equation}
\frac{S l_\nu^2}{\nu} =  \Re_c  \Leftrightarrow l_\nu = \left( \frac{ \Re_c \nu}{S}  \right)^{1/2}  =  \left(\frac{\Re_c}{\Re_L} \right)^{1/2} L\, .
\end{equation}
It is easy to verify that the condition $l_{\rm Z} = l_\nu$ is equivalent to Zahn's stability criterion (\ref{eq:crit2}). When this happens, turbulence disappears and $D_{\rm turb}$ must drop to zero instead of being given by (\ref{eq:zahndt}). 

In the limit of weak stratification on the other hand, it is easy to see from (\ref{eq:zahnscale2}) that the Zahn scale increases as stratification (or equivalently, $J$) decreases. In a numerical experiment with finite domain size $L$, or in real stellar conditions where the shear layer itself has a finite width $L$, $l_{\rm Z}$ must therefore eventually approach $L$ as $J$ decreases, at which point the vertical eddy size becomes limited by the domain size instead of being determined by $l_{\rm Z}$. When this happens, we expect Zahn's mixing prescription to fail. Instead, $D_{\rm turb}$ presumably becomes proportional to $S L^2$ instead of $S l_{\rm Z}^2$. 

To test these ideas, we must compute the height of turbulent eddies $l_e$ and compare it to $L$, $l_{\rm Z}$, and $l_\nu$. From the discussion above, we predict that $l_e$ (1) should be equal to $L$ \citep[or to a fraction of $L$, as proposed by][]{Pratal2016} in very weakly stratified systems where the eddy size is limited by the system size, (2) should be equal to $l_{\rm Z}$ in an intermediate regime, and (3) that turbulence should disappear in strongly stratified systems when $l_e$ approaches $l_\nu$. The computation of $l_e$ is non-trivial, however.  \citet{GaraudKulen16} and \citet{Pratal2016}, who both used models in which perturbations are triply-periodic, estimated the eddy scale using a weighted average of the energy spectrum, as in 
\begin{equation}
l_e =  \left[  \frac{\sum_{\check k_x} \sum_{\check k_y} \sum_{\check k_z \ne 0} E_{\check w}(\check \bk,t) \check k_z^{-1}  }{\sum_{\check k_x} \sum_{\check k_y} \sum_{\check k_z \ne 0} E_{\check w}(\check \bk,t)  } \right]_t \mbox{  ,}
\label{eq:spectralle}
\end{equation}
or variants thereof \citep[see][for detail]{GaraudKulen16,Pratal2016}. In this expression, $E_{\check w}(\check \bk,t)$ is the energy in the Fourier component of wavenumber $\check \bk=(\check k_x,\check k_y,\check k_z)$ of the vertical velocity field. 

\begin{figure}[h]
  \centerline{\includegraphics[width=0.5\textwidth]{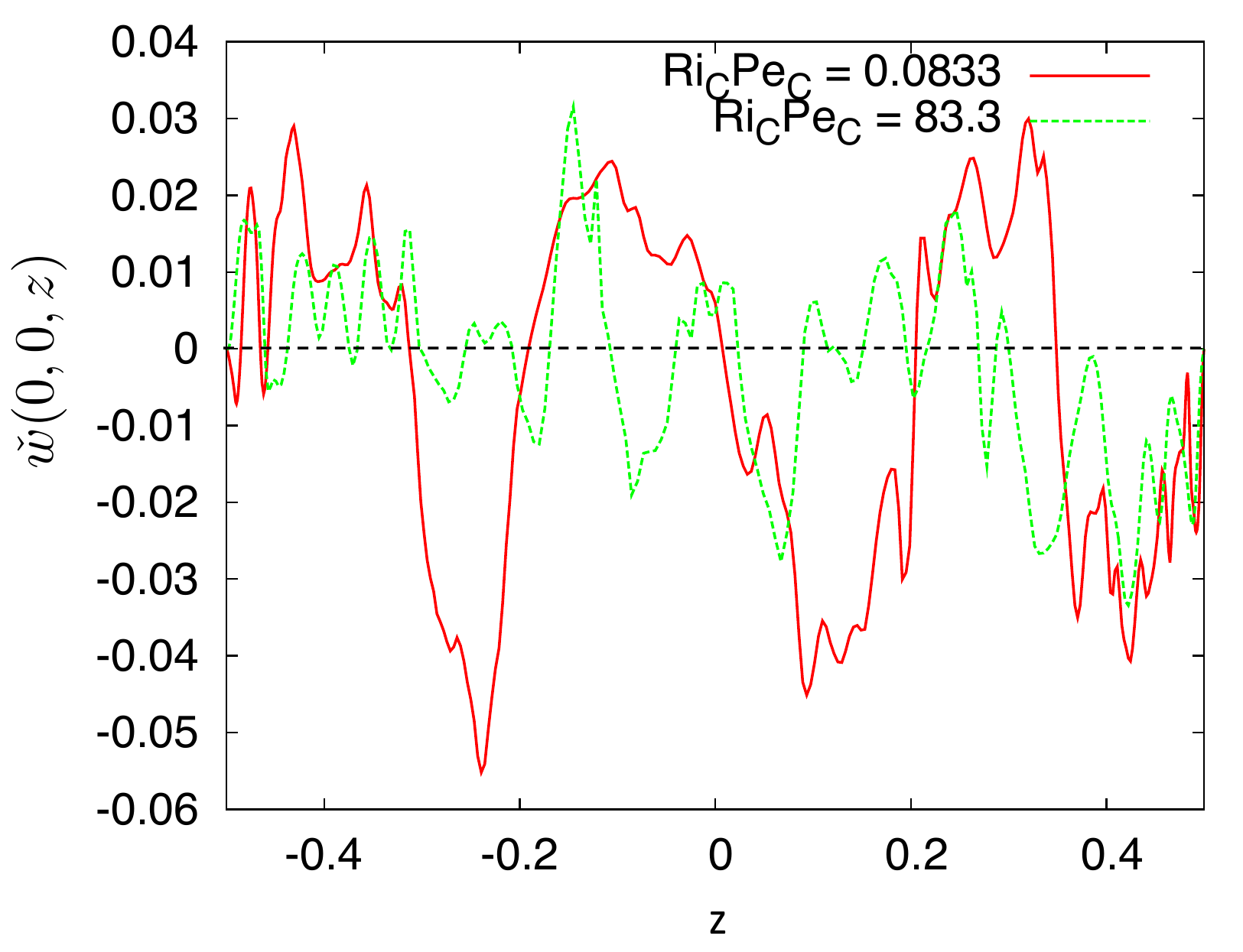}}
  \caption{Profiles of $\check w(0,0,z)$ at randomly selected times during the quasi-stationary state, for $\Re_{\rm C}  = 1.2 \times 10^5$, in the two simulations illustrated in Figure \ref{fig:prettypics}. Note how both profiles contain features on small scales, but only the more weakly stratified one contains features on larger-scales too.}
\label{fig:profiles}
\end{figure}

In a system with impermeable boundaries such as ours, however, this definition cannot be used as is. In any case, as discussed in Section \ref{sec:intromix}, both \citet{GaraudKulen16} and \citet{Pratal2016} found that using this estimate of the eddy scale leads to inconsistencies in the model, namely that while $D_{\rm turb}$ is approximately predicted by Zahn's model for intermediate values of $J$, it is {\it not} proportional to $S l_e^2$ even though this is one of the basic assumptions of Zahn's theory. This statement is equivalent to saying that $D_{\rm turb} = \beta S l_{\rm Z}^2$ even though $D_{\rm turb} \ne \beta S l_e^2$, implying that $l_e \ne l_{\rm Z}$. A possible solution of this conendrum is that the turbulent eddies are indeed of size $l_e = l_{\rm Z}$, but that the definition given in (\ref{eq:spectralle}) is {\it not} a good estimate for $l_e$. This notion is supported by the observation that while the strongly-stratified runs are dominated by small-scale eddies (as expected), the more weakly stratified runs, especially at large Reynolds numbers, contain both small and large scales (see Figure \ref{fig:prettypics} and Figure \ref{fig:profiles}). The Zahn scale, presumably, would correspond to the larger one, but the formula given in (\ref{eq:spectralle}) is strongly weighted towards the smaller scales and would pick these up rather than $l_{\rm Z}$. 

For these reasons, we now propose an alternative way of measuring the eddy scale, using autocorrelation functions of the turbulent flow field. We define 
\begin{eqnarray}
%a_w(l) = \left[  \frac{2}{L-l} \int_0^{L-l} \frac{\check w(0,0,z,t) \check w(0,0,z+l,t)}{\check w^2(0,0,z,t)  + \check w^2(0,0,z+l,t) } dz \right]_t\, ,
a_v(l) = \left[  \frac{1}{L-l} \int_0^{L-l} \check v(0,0,z,t) \check v(0,0,z+l,t) dz \right]_t\, , \label{eq:avl} \\
a_w(l) = \left[  \frac{1}{L-l} \int_0^{L-l} \check w(0,0,z,t) \check w(0,0,z+l,t) dz \right]_t\, \label{eq:awl} ,
\end{eqnarray}
as the mean vertical autocorrelation functions of $\check v$ and $\check w$ respectively. The streamwise velocity $\check u$ is less useful for computing the eddy scale, as it is dominated by the effects of the mean flow even if the latter is subtracted.  Note that we integrate over flow profiles taken at $x=0$ and $y=0$, rather than over the entire flow field, because those are stored at very regular time intervals in our simulations while the whole flow field is only stored once in a while. Since the system is horizontally periodic, and statistically stationary between $t_1$ and $t_2$, we expect that these individual profiles have similar statistics as those of the entire flow; this is actually verified to be true in the Appendix. %Note also that we use this somewhat unusual definition that includes a normalization to avoid the possibility of having very strong individual up- or down-flows dominate the integral. In practice, this provides cleaner data in the limit of weak stratification, and does not affect the limit of strong stratification (where such strong up- and down-flows do not exist anyway). This definition also has the nice property of being equal to $1$ when $ l = 0$. 

\begin{figure}[h!]
  \centerline{\includegraphics[width=0.5\textwidth]{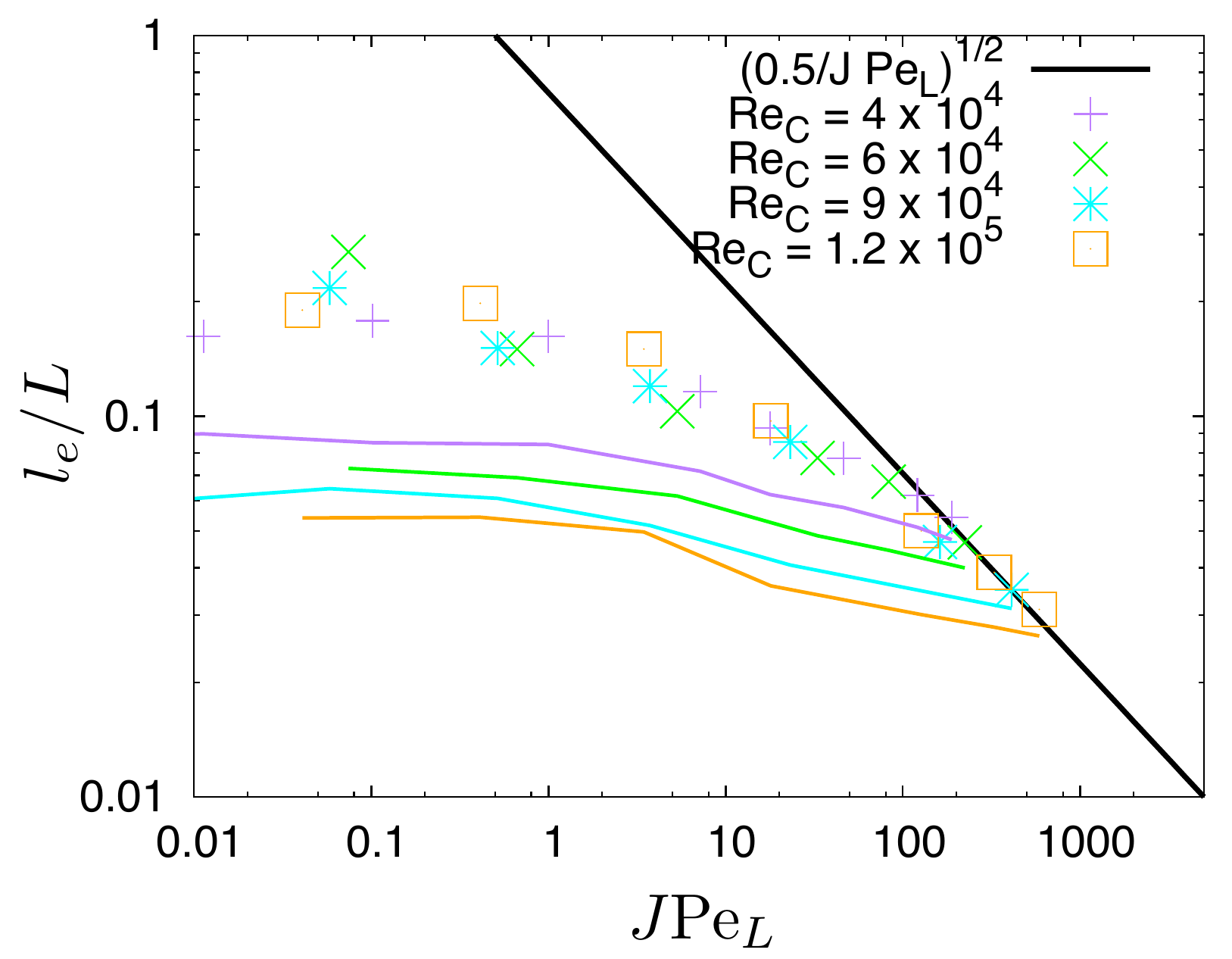}}
  \caption{Comparison between the various theoretical and experimental lengthscales $l_{\rm Z}$ (solid black line), $l_\nu$ (solid colored lines), and $l_e$ (data points) respectively. The constant $(J \Pe)_c$ in the theoretical Zahn scale (see equation \ref{eq:zahnscale2}) has been calibrated to fit the data in the limit of large stratification. That fit then uniquely defines the  constant $\Re_c$ which is needed to compute $l_\nu$. The colors of the lines and symbols are the same for $l_\nu$ and $l_e$ that share the same value of $\Re_{\rm C}$. }
\label{fig:lengthscales}
\end{figure}

There are many possible ways of constructing an eddy lengthscale from these autocorrelation functions. Two of them are explored in the Appendix, and their pros and cons are discussed. We have eventually settled on defining $l_e$ as the first zero of $a_v(l)$:
\begin{equation}
a_v(l_e) = 0.
\label{eq:lev}
\end{equation}
%Note that for this case, very similar results are obtained if we defined $l^{(2)}_e$ as in (\ref{eq:lew}) but using $a_v(l)$ instead of $a_w(l)$.
Figure \ref{fig:lengthscales} shows $l_e/L$ thus measured as a function of $J \Pe_L$ for all the available runs. Generally speaking, we see that $l_e$ decreases with increasing stratification, as expected. In the limit of very large stratification ($J \Pe_L > 100$), all the data for $l_e$ asymptotes to the same line in this log-log graph. Fitting the data, we find that $l_e/L \simeq (0.5/J \Pe_L)^{1/2}$ which in turn shows that $l_e$ is proportional to the Zahn scale, since $l_{\rm Z}/L = ((J\Pe)_c/J \Pe_L)^{1/2}$. In fact, we can set the two equal to one another in this limit, in line with Zahn's model \citep{Zahn92}, which enables us to calibrate the unknown constant $(J\Pe)_c$ to be 
\begin{equation}
(J\Pe)_c \simeq 0.5\, .
\end{equation}
Figure \ref{fig:lengthscales} shows the calibrated Zahn scale in addition to the turbulent eddy scale data. 

Having calibrated $(J\Pe)_c$, we can also determine $\Re_c$ since we had previously found that $(J\Pr)_c = (J \Pe)_c/\Re_c \simeq 0.007$ (see equation \ref{eq:crit2}). This implies 
\begin{equation} 
\Re_c \simeq 71\, .
\end{equation}
Finally, now that $\Re_c$ is known, we can plot $l_\nu$ in Figure \ref{fig:lengthscales}. For each set of runs (i.e. for each value of the input Reynolds number $\Re_{\rm C}$), we see that turbulent solutions only exist up to the point where $l_Z = l_\nu$ but not beyond. This is entirely consistent with the combined models of  \citet{Zahn1974} and \citet{Zahn92}. 

%demonstrate that the ratio of the turbulent eddy size to the global shear scaleheight indeed dictates the transition from one regime to the next, we show in Figure XXXX the vertical scale $l$ of turbulent eddies extracted from a computation of the two-point correlation function of the horizontal velocity $v$ (see Appendix for a discussion of this method), as a function of $J \Pe_L$ (i.e. the Richardson-Peclet number based on the actual shearing rate in the bulk of the simulation, see equation XXX). 
%We see that $l$ is largest in the limit of low stratification, with a size $l \simeq 0.2L$, and decreases progressively as $J\Pe_L$ increases. 

%As shown by \citet{Garaudal15}, a stratified shear flow is energy stable in the limit where $J \Pr$ exceeds a certain threshold of order unity, so that $D_{\rm turb}$ must be identically 0 beyond that threshold. As found by \citet{Pratal2016}, and confirmed in Figure 
%\ref{fig:DtvsJPr}, stratified shear flows are in fact stable somewhat earlier, around $J \Pr \simeq 0.007$. This explains the sharp drop of the function $D_{\rm turb}J/\kappa_T$ near this critical value. In the opposite limit, as $J \Pr \rightarrow 0$, one should also expect the model given in Equation (\ref{eq:zahndt}) to break down. Indeed, $D_{\rm turb}$ would tend to infinity in the strict limit of $J \rightarrow 0$, which is clearly unphysical. One might expect that $D_{\rm turb}$ should instead tend to a finite value that models turbulent transport in an unstratified fluid. 

Looking at progressively more weakly stratified systems, Figure \ref{fig:lengthscales} shows that $l_{\rm Z}$ stops being a good estimate for the eddy scale $l_e$ when $J \Pe_L$ decreases below $\sim 100$. Instead, $l_e$ tends to a (roughly) constant fraction of the total domain height (about $20\%$) when $J\Pe_L \rightarrow 0$. This confirms our suspicion that the local approximation made by \citet{Zahn92} must fail at low stratification \citep[see also][]{Pratal2016}. As discussed earlier, a better model in this limit would be one in which, for instance, $D_{\rm turb}$ becomes proportional to $SL^2$ rather than $Sl_{\rm Z}^2$. 

\begin{figure}[h!]
  \centerline{\includegraphics[width=0.5\textwidth]{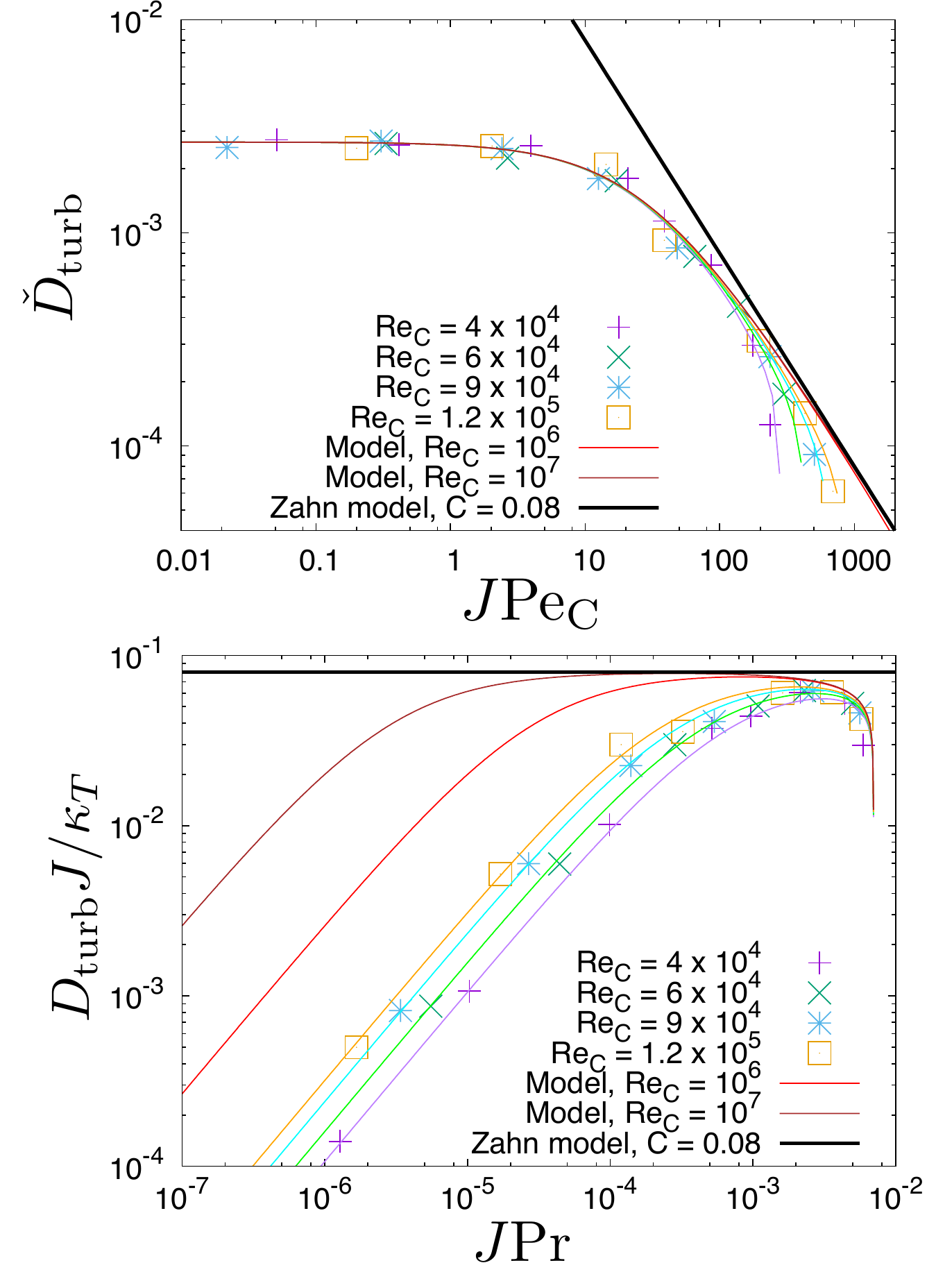}}
  \caption{Comparison of the data and the model for the turbulent diffusivity $D_{\rm turb}$. The top panel shows this comparison in terms of the non-dimensional diffusivity $\check D_{\rm turb}$ vs. $J \Pe_{\rm C}$, which emphasizes the behavior of the system in the weakly stratified limit. The bottom panel shows the same data and models in terms of $D_{\rm turb} J/\kappa_T$ vs. $J \Pr$, which emphasizes the behavior of the system in the strongly stratified limit. In both panels the symbols show the data, and the lines of the corresponding color show the model predictions for the same $\Re_{\rm C}$. Also shown are model predictions for $\Re_{\rm C} = 10^6$ and $\Re_{\rm C} = 10^7$. Note that the errorbars have been omitted to avoid crowding the plot. The thick black line represents Zahn's basic model (\ref{eq:zahndt}) with $C = 0.08$.}
\label{fig:newmodel}
\end{figure}

In fact, such a model can be obtained using simple dimensional analysis. Indeed, the limit  $J \rightarrow 0$ is equivalent to the limit $\Ri_{\rm C} \rightarrow 0$ (and therefore $\Ri_{\rm C}\Pe_{\rm C} \rightarrow 0$) since the two are related via (\ref{eq:Jdef1}). As a result the non-dimensional $\check{D}_{\rm turb}$ should asymptote to a value that only depends on the Reynolds number, since it is the only remaining non-dimensional parameter in that limit. If, furthermore, we recall that $\Re_{\rm C} \gg 1$ in stellar interiors, we can anticipate that $\check{D}_{\rm turb}$ should tend to a constant $\check{D}_0$ that is independent of $\Re_{\rm C}$ as well\footnote{Note that $\check D_0$ is model-specific, in the sense that using a different model setup would yield a different constant.}. We see in Figure \ref{fig:newmodel}a, which shows $\check{D}_{\rm turb}$ as a function of $J \Pe_{\rm C}$, that this is indeed the case even for moderate Reynolds numbers, and estimate that $\check{D}_0 \simeq 0.025$. Dimensionally speaking, this implies $D_{\rm turb} \simeq \check{D}_0 S_{\rm C} L^2$, and since $S \simeq 0.2 S_{\rm C}$ in that limit (see Figure \ref{fig:Jcompare}), we recover $D_{\rm turb} \propto S L^2$ as expected. 

In terms of the quantity $D_{\rm turb} J /\kappa_T =  \check{D}_{\rm turb} J \Pe_{\rm C}$, we then have 
\begin{equation}
\lim_{J \rightarrow 0} \frac{D_{\rm turb} J }{ \kappa_T} = \check{D}_0 J \Pe_{\rm C} = \check{D}_0 J \Pr \Re_{\rm C}\, ,
\label{eq:lowJlimit}
\end{equation}
which is clearly quite different from the behavior of Zahn's model or of the parametric fit of \citet{Pratal2016} in the same limit (where $D_{\rm turb}J/\kappa_T$ tends to a constant and has no dependence on $\Re_{\rm C}$). In Figure \ref{fig:newmodel}b, we see that our data indeed supports the notion that $D_{\rm turb} J/ \kappa_T \propto J \Pr \Re_{\rm C}$ in the limit of small $J\Pr$. 

\subsection{A new model}

We now set out to propose a new model that correctly captures all three limits: the limit of low $J$, where $D_{\rm turb}J /\kappa_T$ should be a function $J \Pe_{\rm C}$ by dimensional analysis, the limit of large $J$, where $D_{\rm turb}J /\kappa_T$ should drop to zero when $J \Pr \rightarrow (J\Pr)_c$ and the intermediate regime where Zahn's local model \citep{Zahn92} directly applies and $D_{\rm turb}J /\kappa_T$ should tend to a constant. This lead us to propose the following functional form:
\begin{equation}
\check{D}_{\rm turb} = f_1( J \Pe_{\rm C}) f_2(J \Pr)\, ,
\end{equation}
where the functions $f_1$ and $f_2$ satisfy the following properties: (1) $f_2$ tends to 1 for small $J \Pr$, (2) $f_2$ tends to 0 as $J \Pr$ tends to $(J \Pr)_c \simeq 0.007$, (3) $f_1$ tends to about $\check D_0 \simeq 0.025$ as $J \Pe_{\rm C}$ tends to 0, (4) $f_1$ tends to $C (J \Pe_{\rm C})^{-1}$ for large $J \Pe_{\rm C}$ (but small $J \Pr$). The first condition is simply a normalization condition. The second is to satisfy the empirical stability criterion. The third is to match the data in the limit of small $J \Pe_{\rm C}$ and the fourth, finally, is to recover Zahn's model in the intermediate limit where $J \Pe_{\rm C}$ is large but $J \Pr$ is small. After some experimentation, we have found that the following model matches the data quite well, and is reasonably simple:
\begin{eqnarray}
&& \check{D}_{\rm turb} = \frac{C}{a+J \Pe_{\rm C}} \left( 1 - \frac{J \Pr}{(J\Pr)_c} \right)^{b}\, , \mbox{  if   } J\Pr < (J\Pr)_c\, , \nonumber \\
&& \check{D}_{\rm turb} = 0 \mbox{  if   } J\Pr > (J\Pr)_c\, ,
\end{eqnarray}
or equivalently
\begin{eqnarray}
&& \frac{D_{\rm turb} J}{\kappa_T} = \frac{C}{1 + a(J \Pe_{\rm C})^{-1} } \left( 1 - \frac{J \Pr}{(J\Pr)_c} \right)^{b}\,  \mbox{  if   } J\Pr < (J\Pr)_c\, , \nonumber \\
&& \frac{D_{\rm turb} J}{\kappa_T}  = 0 \mbox{  if   } J\Pr > (J\Pr)_c\, .
\label{eq:ourmodel}
\end{eqnarray}
Note that $C / a = \check D_0 \simeq 0.025$, so the constants $C$ and $a$ are not independent of one another. Fitting the data, we find that using
\begin{equation}
C \simeq 0.08\, , a \simeq 30\, , b \simeq 0.25, \mbox{  and  } (J\Pr)_c \simeq 0.007
\end{equation}
provides a reasonably good overall fit to all of our simulations\footnote{Note that until more data is obtained in the large Reynolds number limit to improve our constraints on $C$, it is somewhat pointless to attempt to fit these constants within high degrees of accuracy.}. Note that having found $C$, we can then obtain the constant $\beta$ of Zahn's model (see equation \ref{eq:dturblz}): since $C = \beta (J\Pe)_c$ (see equation \ref{eq:zahndt}) we have
\begin{equation}
\beta \simeq 0.16 \, .
\end{equation}

The model is shown in Figure \ref{fig:newmodel}, for various values of the Reynolds number corresponding to the numerical simulations available, as well as for much larger Reynolds numbers for which simulations are not currently possible, namely $\Re_{\rm C} = 10^6$ and $\Re_{\rm C} = 10^7$. 

In the limit of very high Reynolds numbers ($\Re_{\rm C} = 10^7$), the model prediction clearly exhibits the three significant regimes discussed above: the very high $J$ cutoff, in which the model only depends on $J \Pr$ and where $D_{\rm turb}$ drops to zero, the very low $J$ limit where $\check{D}_{\rm turb} \rightarrow \check{D}_0$ or equivalently $D_{\rm turb}J/\kappa_T \rightarrow \check{D}_0 J \Pr \Re_C$, and an intermediate region where $J \Pe_{\rm C}$ is large while $J \Pr = J \Pe_{\rm C} / \Re_C$ is small, where $D_{\rm turb}J/\kappa_T \simeq C$, or equivalently $\check{D}_{\rm turb} \simeq  C(J \Pe_{\rm C})^{-1}$. As the Reynolds number decreases, however, the intermediate region shrinks and the dynamics rapidly go from being domain-size dominated to viscously dominated. In the $\Re_{\rm C} = 10^6$ case, for instance, one only barely distinguishes the regime where $D_{\rm turb}J/\kappa_T \simeq C$. This explains why in our numerical simulations \citep[and similarly in those of][]{Pratal2016}, where the largest value of $\Re_{\rm C}$ is of the order of $10^5$, the intermediate region is effectively absent. Despite this, we see that the new model fits the data very well for all available simulations, both in the strongly stratified limit of course, where it recovers the essence of Zahn's 1974 and 1992 models, but also in the weakly stratified limit where it accounts for the limitation of the eddy scale either by the domain size or by the shear lengthscale. 

\subsection{Discussion about the proposed model}
\label{sec:modeldisc}

The proposed model and its fitted constants, namely those used in (\ref{eq:ourmodel}) as well as the others introduced in Section \ref{sec:intro} ($\Re_c$, $(J\Pe)_c$ and $\beta$) require some discussion. 

First, note that $C$, $b$ and $(J\Pr)_c$ are tied to the properties of the local model only -- the first being from Zahn's original model (\ref{eq:zahndt}), and the other two being related to the proposed correction in the limit of $J \Pr \rightarrow (J\Pr)_c$. In this sense, they should be fairly universal and are not expected to depend, say, on the large-scale properties of the shear or on the boundary conditions applied in the model. The constraints on $(J\Pr)_c$ are fairly strong given that values close to $0.007$ were independently measured in various model setups by \citet{Pratal2016}, \citet{GaraudKulen16} and in this paper. Strict errorbars on $(J\Pr)_c$ still remain to be estimated from a larger suite of simulations however. The constraints on $b$ are arguably weaker. In fact, whether the variation of $D_{\rm turb}J/\kappa_T$ in the limit of $J \Pr \rightarrow (J\Pr)_c$ is best represented by a power law, as we propose here, or by another function that tends to zero at $J \Pr = (J\Pr)_c$, remains to be determined. In the meantime, the proposed power law with $b \simeq 0.25$ seems to be adequate. Finally, it is important to note that the model constraints on $C$ are also quite weak at this point. The constant $C$ uniquely controls the value of $D_{\rm turb}$ in the intermediate regime where the Zahn scale is much smaller than the domain size, and yet much larger than the viscous scale. Unfortunately, this regime is never achieved with the values of the Reynolds number presently accessible to moderate-scale numerical simulations. Instead, we are forced to fit $C$ under less-than-ideal conditions where it {\it partially} contributes to the variation of $D_{\rm turb}$ in the limits where $J\Pr \rightarrow (J\Pr)_c$ or $J\Pr \rightarrow 0$, together with all the other model constants. Hence uncertainties in estimating $a$, $(J\Pr)_c$ and $b$ all affect our estimate for $C$. To address this problem, we plan to run a few simulations at much larger Reynolds number ($\Re_{\rm C} = O(10^6)$) in the future to better constraint $C$ independently of the other constants.

Second, it is crucial to understand that, by contrast with $C$, $b$ and $(J\Pr)_c$, the constant $a$ is {\it not} universal but instead depends on the global  properties of the model (such as the shape and amplitude of the mean shear, its lengthscale compared with the domain size, and the manner in which the model is non-dimensionalized). To see this more clearly, consider a thought experiment where we simply non-dimensionalize our plane Couette flow model in two different ways, using, say, the original lengthscale $L$ in one case and $L/2$ in the other. The prediction for $D_{\rm turb}J/\kappa_T$ must remain unchanged, since the (dimensional) model setup is the same in both cases. However since the P\'eclet number definitions are different (one being $\Pe_1 = \Pe_{\rm C}$ and the other being $\Pe_2 = \Pe_{\rm C}/4$), the value of $a$ would have to change by a factor of 4 to compensate (see equation \ref{eq:ourmodel}). The constant $a$ would similarly have to change should one decide to use a different unit time than $S_{\rm C}$. In other words, $a$ is inherently model dependent.  

\begin{figure}[h]
  \centerline{\includegraphics[width=0.5\textwidth]{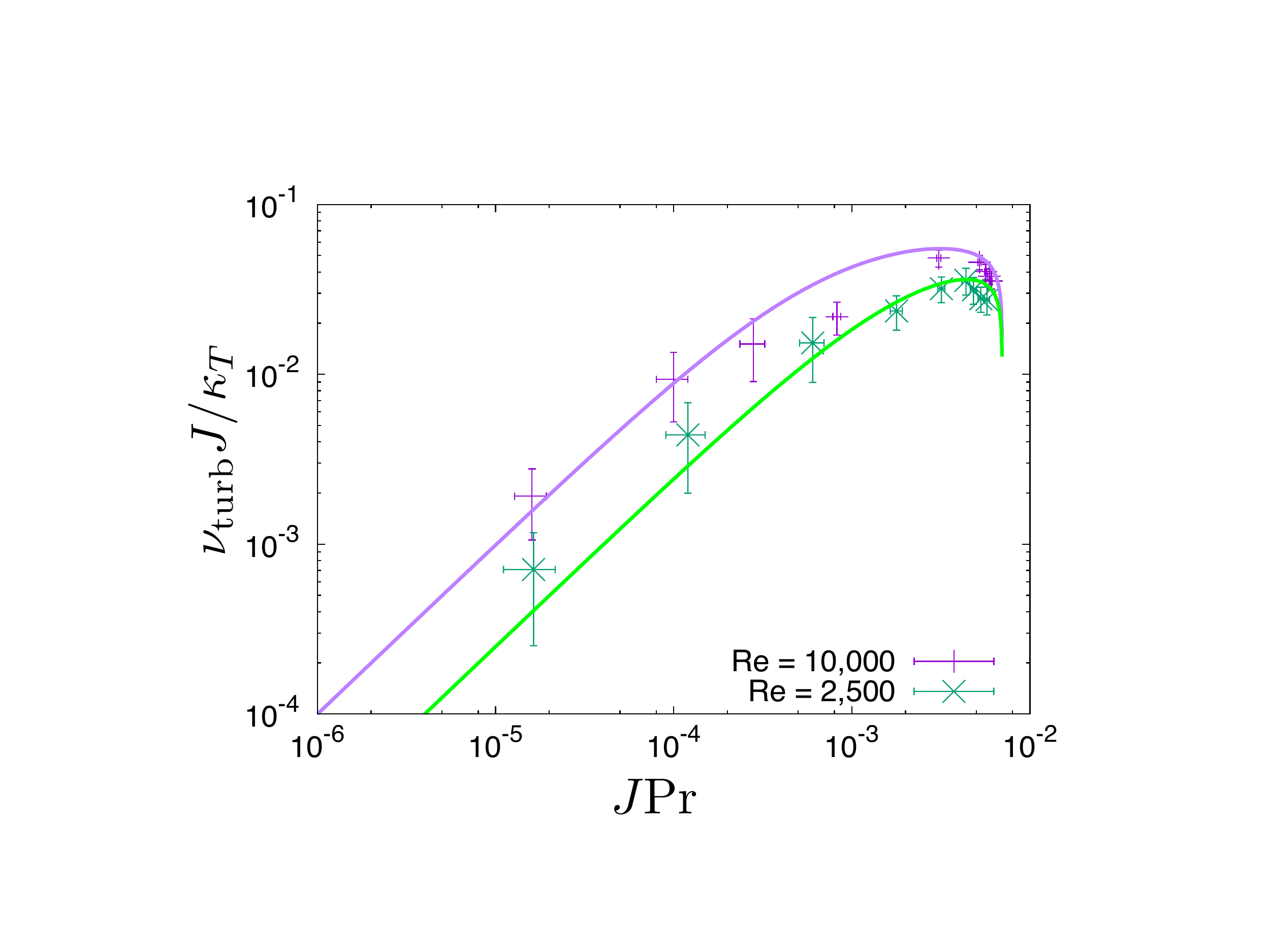}}
  \caption{Comparison of our new model (see equation \ref{eq:ourmodello}) with the data from body-forced sinusoidal shear profiles of \citet{GaraudKulen16}. The constants $C = 0.08$, $b =0.25$, $(J\Pr)_c = 0.007$ have not been changed, but $a$ has been re-fitted and found to be about 8 in this case.} 
\label{fig:GaraudKulen}
\end{figure}

Hence, if one wishes to apply equation (\ref{eq:ourmodel}) to predict how much mixing will take place in a system different from the one introduced in Section \ref{sec:model}, $C$, $b$ and $(J\Pr)_c$ would remain unchanged but $a$ needs to be re-calibrated. To verify that this is indeed the case, and by way of providing an example, we revisit the data from \citet{GaraudKulen16} for the sinusoidally body-forced shear flow (see Section \ref{sec:intro}). We extract the mean shear $S$ at $z = \pi$ (which is in the middle of the domain and in the middle of the turbulent shear layer in their simulations), as well as the turbulent viscosity\footnote{Note that \citet{GaraudKulen16} did not add a passive scalar field for their low P\'eclet number runs, which is why we use here the turbulent viscosity instead. } at that position using the technique described in this paper, see Section \ref{sec:mixing}. We then compare $\nu_{\rm turb}J/\kappa_T$ to the model predictions, this time using the formula
\begin{equation}
\frac{\nu_{\rm turb} J}{\kappa_T} = \frac{C}{1 + a(J \Pe)^{-1}} \left( 1 - \frac{J \Pr}{(J\Pr)_c} \right)^{b}\,   \mbox{  for  } J\Pr < (J\Pr)_c\, ,
\label{eq:ourmodello}
\end{equation}
where $\Pe$ is the input P\'eclet number based on the laminar flow \citep[see the definition in][]{GaraudKulen16}, where $C = 0.08$, $b =0.25$, $(J\Pr)_c = 0.007$, but where $a$ has been re-fitted and found to be about 8. Note that we have assumed for simplicity that $\nu_{\rm turb} \simeq D_{\rm turb}$. The results are shown in Figure \ref{fig:GaraudKulen}. We see that, with this re-fitting, the model adequately captures nearly all the features of the data from the weakly to the strongly stratified limit, giving us confidence that it is indeed the correct approach and that our model is fairly universal (aside from $a$). 

Of course, it is a little disappointing -- but unavoidable given its meaning -- that there should be a model-dependent constant left. The value of $a$ can be determined from numerical experiments under a specific model setup, but it remains unclear what to select for it in the context of a stellar evolution model where the general shape of the shear, and the forcing mechanisms, vary from star to star. In practice, however, the uncertainty in $a$ should not affect the model predictions too much in the limit of large Reynolds numbers appropriate for stellar interiors. Indeed, as shown in Figure \ref{fig:newmodel}, the region of parameter space where the value of $a$ influences $D_{\rm turb}J/\kappa_T$ is limited to that of weak stratification where $ D_{\rm turb}J/\kappa_T \simeq (C/a) J \Pe_{\rm C}$, and the size of that region shrinks with increasing Reynolds number. One may therefore hope that the values of $J \Pr$ actually relevant for stars would always fall in either the intermediate regime, or in the strongly stratified regime. In the meantime, for the purpose of picking a value of $a$ to be used in stellar evolution codes, one might as well use something completely generic such as $a = 1$. 

Finally, a related item of discussion concerns the model constants $(J\Pe)_c$, $\Re_c$ and $\beta$.  By contrast with $C$, $b$ and $(J\Pr)_c$, they are not universal because they depend on the definitions we have used to evaluate the eddy scale $l_e$. To see this, suppose we had instead measured the vertical eddy scale using the autocorrelation function of the vertical velocity field $a_w(l)$ instead of $a_v(l)$ (see equation \ref{eq:awl} and the Appendix for detail). Applying the same steps as in Section \ref{sec:lengthscales}, we would have found that  $(J\Pe)_c$ is roughly equal to $2$ instead of 0.5, that $\Re_c \simeq 284$, and $\beta = 0.04$. One should therefore always bear the relative arbitrariness of $(J\Pe)_c$, $\Re_c$ and $\beta$ in mind. Ultimately however, this does not pose any problem since our model has been written in such a way as to contain only the constants $C$, $b$ and $(J\Pr)_c$ that are indeed universal, together with $a$ which was discussed earlier.  

%To summarize, we find that 
%\begin{equation}
%\frac{D_{\rm turb} J}{\kappa_T} = \frac{0.08 }{1 + b(J \Pe_C)^{-1} } \left( 1 - \frac{J \Pr}{0.007} \right)^{1/4}
%\end{equation}
%where the constants $0.08$, and $0.007$, are universal. On the other hand, teh constant $b$ is not universal and depends on the large-scale structure of the unstratified shear flow (the size of the layer, and the shear within the layer). 

\section{Summary and future prospects}
\label{sec:ccl}

In this paper we have analyzed a suite of numerical experiments of diffusive stratified shear instabilities using a plane Couette flow setup, and under both the Boussinesq approximation \citep{SpiegelVeronis1960} and the low P\'eclet number approximation \citep{Lignieres1999}. We were able to span a range of Reynolds numbers from $4\times 10^4$ up to $1.2 \times 10^5$, and a range of Richardson-P\'eclet number from about $10^{-4}$ to about $10^4$. Our results are very comparable with those of Prat and collaborators \citep{PratLignieres13,PratLignieres14,Pratal2016} even though they used a different model setup, and the favorable comparison can be viewed as a successful validation of the codes and of our selected approaches. 

We have analyzed our results in the light of Zahn's models \citep{Zahn1974,Zahn92} and found that (1) the stability criterion of \citet{Zahn1974}, given in (\ref{eq:crit2}),  appropriately describes the upper limit for turbulent mixing in simulations of stratified shear flows, with $(J\Pr)_c \simeq 0.007$ as initially found by \citet{Pratal2016}; and (2) the turbulent mixing model of \citet{Zahn92} given in (\ref{eq:zahndt}), on the other hand, needs to be augmented to account for the stability cutoff in the strongly stratified limit, and for non-local effects in the weakly stratified limit. For intermediate values of the stratification, on the other hand, we have provided some evidence that the model is likely appropriate, especially at large Reynolds numbers.

In order to better account for all three regimes (weak, intermediate and strong stratification), we then proposed a new model for diffusive stratified shear instabilities that extends and unifies the models of \citet{Zahn1974} and \citet{Zahn92} and fits the results of our numerical experiments quite well. In this model, the turbulent diffusivity (of a passive tracer) is given by 
\begin{equation}
D_{\rm turb} = \frac{C}{1 + a(J \Pe)^{-1}} \left( 1 - \frac{J \Pr}{(J\Pr)_c} \right)^{b} \frac{\kappa_T}{J} \mbox{  if   } J\Pr < (J\Pr)_c\, ,
\label{eq:ourmodelfinal}
\end{equation}
where $\Pe = SL^2/\kappa_T$ and $L$ is the shear scaleheight. The model constants, which have been fitted to our data, include 
\begin{equation}
C \simeq 0.08\, , b \simeq 0.25, \mbox{  and  } (J\Pr)_c \simeq 0.007
\end{equation}
which are universal constants, and $a$ which depends on the model setup considered. Absent further information, we suggest the use of $a = 1$ in stellar evolution calculations just for simplicity. While $a$ plays an important role in the model predictions in numerical simulations at moderate Reynolds numbers, it should not really affect the predictions for $D_{\rm turb} J/\kappa_T$ 
in the limit of large Reynolds numbers and large $J$ appropriate of stellar interiors (see Section \ref{sec:modeldisc} for a discussion of this issue). Finally, we have also shown that 
\begin{equation}
\nu_{\rm turb} \simeq \gamma D_{\rm turb}  \, ,
\end{equation}
where the proportionality constant $\gamma$ appears to vary between about 0.8 and 1 depending on the regime considered. This result is similar to the findings of \citet{Pratal2016}. Since this variation could be due to the influence of boundaries in our system, we prefer not to attempt to constrain it further here. Nevertheless, if a simple order-of-magnitude estimate is required, selecting $\gamma = 1$ is appropriate. 

As discussed in Section \ref{sec:modeldisc}, while $(J\Pr)_c$ is reasonably well constrained, significant uncertainties remain on $b$ and $C$. Uncertainties on $b$ can be reduced by running and analyzing a suite of numerical simulations that systematically explore the region of parameter space where $J \Pr \rightarrow (J\Pr)_c$. Uncertainties on $C$ can be reduced by running and analyzing a few numerical simulations at significantly higher values of the Reynolds number than what we have presented here. In both cases, the tasks requires significant computational resources, but will be pursued in the future. 

In addition to the uncertainties on the model constants, three additional issues remain to be studied. The first is whether Zahn's models \citep{Zahn1974,Zahn92} and their extensions discussed in this paper, which were principally derived and tested under conditions where the shear is constant and where the entire computational domain is turbulent, would also apply in stellar situations where turbulent regions would likely coexist with stable regions, and where the shear may be quite far from being constant. \citet{GaraudKulen16} used a sinusoidal body-force that generated a non-constant mean shear $S(z)$, but focused on analyzing the global properties of their simulations, namely the volume-averaged turbulent diffusion coefficient, and whether the system as a whole becomes turbulent or not. It would be interesting to revisit their simulations to see whether the new model given in (\ref{eq:ourmodelfinal}) applies locally as well (i.e at each vertical position $z$). The preliminary investigation presented in Figure \ref{fig:GaraudKulen} suggests that it might, at least deep in the middle of the turbulent layer. Whether this continues to be true in all circumstances, especially in turbulent regions bordering stable ones, remains to be determined and will be the subject of Paper 2 in this series. 

The second issue concerns the effect of rotation. Zahn's models \citep{Zahn1974,Zahn92} are principally invoked in the context of rotational shear, and in fact the expression for the turbulent diffusion coefficient is often written as 
\begin{equation}
D_{\rm turb} = C \frac{\kappa_T}{N^2} \left( r \frac{d\Omega}{dr} \right)^2\, ,
\end{equation}
where the shear has been implicitly derived from the rotational shear $r d\Omega/dr$ (where $\Omega(r)$ is the rotation rate of a shell at radius $r$ in the star). However, rotation was not included in any of the simulations performed to date. This is problematic because rotation can have both stabilizing and destabilizing effects on stratified fluids, depending on the sign of the angular momentum gradient, which could dramatically change the results obtained so far. Hence, future work must include rotation to provide a more comprehensive test of the theories proposed by Zahn, and in this paper. 

Finally the third issue concerns the limit of validity of the model beyond the use of the low P\'eclet number approximation. In this paper, we have restricted our attention to numerical simulations that use the LPN equations \citep{Lignieres1999}. As discussed by \citet{Lignieres1999}, and confirmed in the numerical simulations of \citet{GaraudKulen16}, this approximation is valid as long as the {\it turbulent} P\'eclet number of the fluid is smaller than one, but not when it exceeds one. Whether Zahn's models would continue to apply when the LPN approximation is invalid remains to be determined. This is a rather formidable computational problem since it requires having a large P\'eclet number together with a small Prandtl number, which implies a {\it very} large Reynolds number. Nevertheless it is a crucial problem since, in all likelihood, a significant fraction of stellar shear layers may actually belong to this region of parameter space rather than to the low P\'eclet number limit. 

\acknowledgements 

P.G. and D.G. gratefully acknowledge funding by NSF AST-1517927. Support was provided for J.V. by the National Aeronautics and Space Administration 
under grants OPR NNX13AK94G and PGG NNX14AN70G. The simulations were run on the Hyades supercomputer at UCSC, purchased using an NSF MRI grant.

\appendix 
\section*{Appendix A}

In this appendix we study the autocorrelation functions $a_v(l)$ and $a_w(l)$ defined in equations (\ref{eq:avl}) and (\ref{eq:awl}) and the manner in which they can be used to define a vertical eddy lengthscale. 

We begin by comparing the autocorrelation functions $a_v(l)$ and $a_w(l)$ with 
\begin{eqnarray}
%a_w(l) = \left[  \frac{2}{L-l} \int_0^{L-l} \frac{\check w(0,0,z,t) \check w(0,0,z+l,t)}{\check w^2(0,0,z,t)  + \check w^2(0,0,z+l,t) } dz \right]_t\, ,
A_v(l,t) =   \frac{1}{L^2} \iint \left[  \frac{1}{L-l} \int_0^{L-l} \check v(x,y,z,t) \check v(x,y,z+l,t) dz \right] dxdy\, , \label{eq:bigavl} \\
A_w(l,t) =   \frac{1}{L^2} \iint \left[  \frac{1}{L-l} \int_0^{L-l} \check w(x,y,z,t) \check w(x,y,z+l,t) dz \right] dxdy\, , \label{eq:bigawl}
\end{eqnarray}
namely the vertical autocorrelation functions at individual times $t$, but integrated over the entire domain. 

In theory, the best way of computing the mean vertical autocorrelation of the flow field would be to take the time-average of the functions $A_v(l,t)$ and $A_w(l,t)$ once the turbulent shear flow is in a statistically stationary state. In practice, however, the functions $A_v(l,t)$ and $A_w(l,t)$ take significantly longer to compute than $a_v(l)$ and $a_w(l)$, and require storage of the entire flow field at each time $t$, which we only have for a handful of simulations. Computing $a_v(l)$ and $a_w(l)$ on the other hand only requires storing single vertical profiles at $x = y = 0$ at each time $t$, which is significantly less expensive and was done by default in all the runs. However, one may wonder whether the functions $a_v(l)$ and $a_w(l)$ computed this way are statistically equivalent to the time-averages of $A_v$ and $A_w$ or not. 

\begin{figure}[h]
  \centerline{\includegraphics[width=\textwidth]{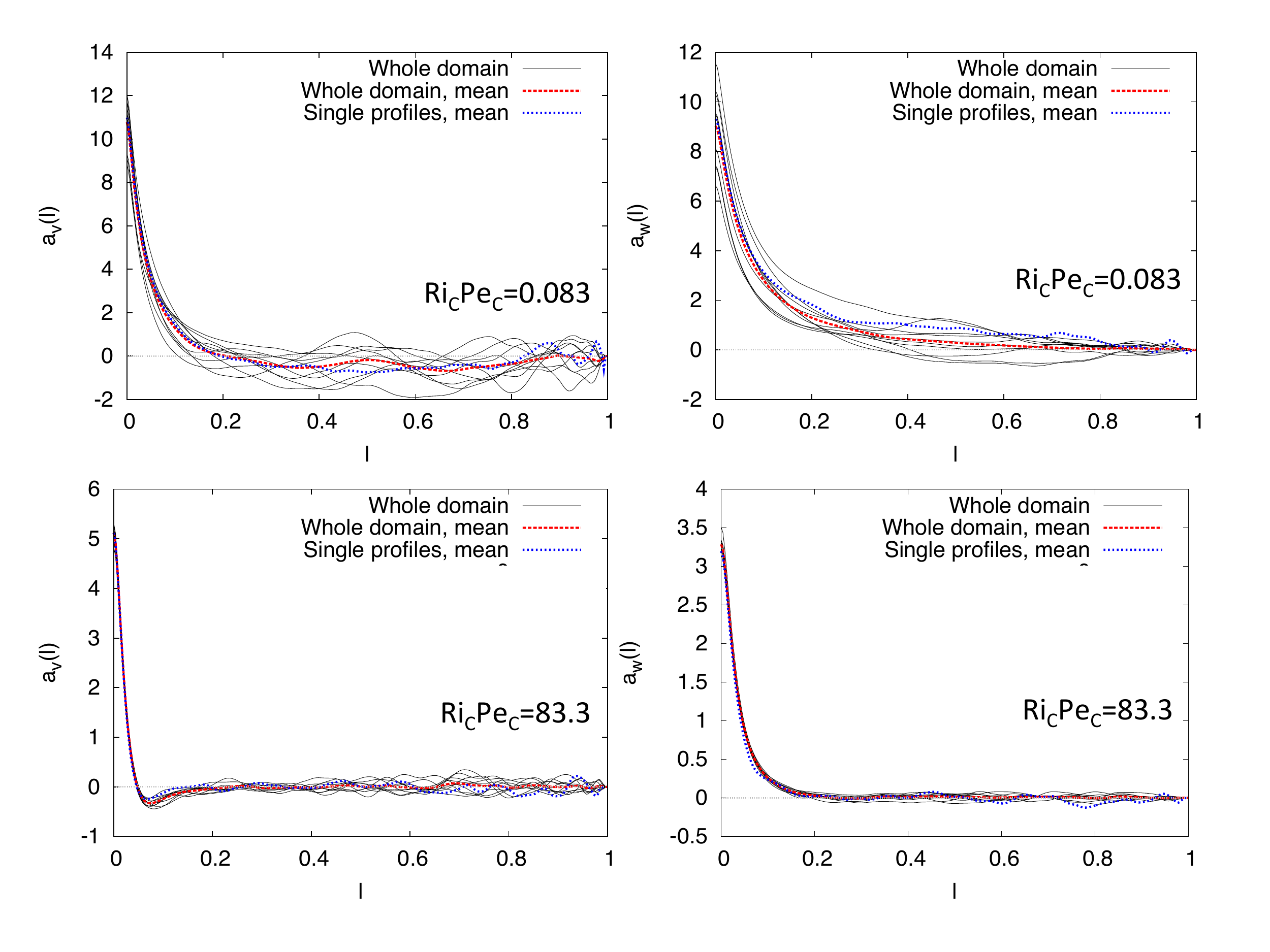}}
  \caption{Comparison of $A_v(l,t)$ and $A_w(l,t)$ at 10 individual timesteps (labeled ``Whole domain", shown in thin black lines), with their average (labeled ``Whole domain, mean", shown in the red dashed line), and with $a_v(l)$ and $a_w(l)$ (labeled ``Single profiles, mean", shown in the blue dotted line), for $\Re_{\rm C} = 1.2 \times 10^5$, in the two simulations illustrated in Figure \ref{fig:prettypics} with $\Ri_{\rm C} \Pe_{\rm C} = 0.083$ and 83.3 respectively.  }
\label{fig:autocorel}
\end{figure}

To test this, we have computed $A_v(l,t)$ and $A_w(l,t)$ at 10 roughly equidistant individual timesteps at which the whole flow field was saved, in the two simulations presented in Figure \ref{fig:prettypics}, as well as the algebraic average of $A_v(l,t)$ and $A_w(l,t)$ over these 10 steps. This is of course not the same as taking the time average of $A_v(l,t)$ and $A_w(l,t)$, but is at least an attempt at approximating that time-average. We then compare in Figure \ref{fig:autocorel} both individual profiles of $A_v(l,t)$ and $A_w(l,t)$ and their averages to $a_v(l)$ and $a_w(l)$ computed for the same simulations.

Figure \ref{fig:autocorel} reveals a number of things. First of all, we see that $a_v(l)$ and $a_w(l)$ (dotted blue lines) are indeed a good approximation to the true vertical autocorrelation function of the flow field (approximated here by the dashed red curves) for low and high stratification, at least for values of $l$ smaller than about 0.5. This confirms that using single profiles at $x=y=0$ to compute the vertical autocorrelation functions of the entire flow field is indeed satisfactory.

Second, looking at $a_v(l)$ and $a_w(l)$ in more detail, we see that they are much more variable, and drop much more slowly with $l$ in the weakly stratified case than in the strongly stratified case. These results are expected from the notion that vertical motion becomes progressively more restricted as stratification increases. Somewhat more surprising is the fact that $a_w(l)$ has a systematically longer ``tail" at large $l$ than $a_v(l)$ at the same value of the stratification, and does not necessarily change sign while $a_v(l)$ always does. We believe that this is because we are comparing somewhat different ideas, namely the {\it vertical} correlation of {\it vertical} flows (in $a_w(l)$) to the {\it vertical} correlation of {\it horizontal} flows (in $a_v(l)$) -- the former is indeed likely to extend further than the latter.  

Naively speaking, the most obvious way of defining the eddy lengthscale $l_e$ would be to take the first zero of $a_w(l)$, interpreting this as the scale over which vertical fluid motions change from being upward to downward, or vice-versa. In practice, however, Figure \ref{fig:autocorel} shows that this definition would be problematic since $a_w(l)$ does not always reliably have a zero. One possible way to solve the problem is to define $l_e$ instead as the value of $l$ where $a_w(l)$ first drops below a certain fraction, say $5\%$, of $a_w(0)$: 
\begin{equation}
a_w(l_e) = 0.05a_w(0) \, .
\label{eq:lew}
\end{equation}
The value of $5\%$ selected here is somewhat arbitrary, in as much as we have found that any value between $5\%$ and $10\%$ yields predictions for $l_e$ that are qualitatively similar. Selecting smaller values picks up too much of the statistical noise, while selecting larger values may not appropriately define an eddy size. As another possibility, having noticed that $a_v(l)$ consistently changes sign at a given lengthscale $l$, we can also define $l_e$ as the lengthscale for which 
\begin{equation}
a_v(l_e) = 0\ \, .
\label{eq:levapp}
\end{equation}

\begin{figure}[h]
  \centerline{\includegraphics[width=0.5\textwidth]{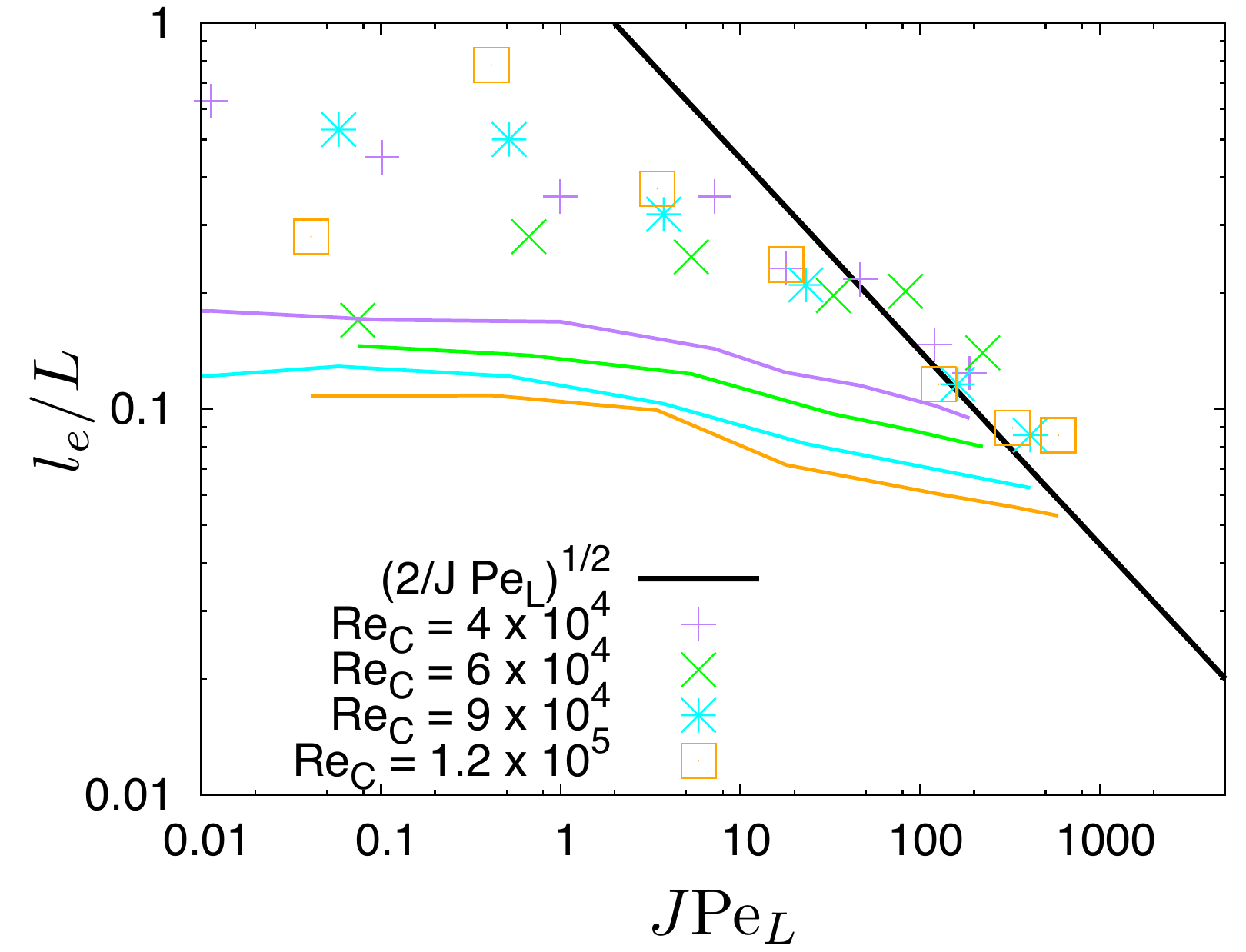}}
  \caption{Estimate of the vertical eddy size using equation (\ref{eq:lew}), for all of the simulations presented in Table 1. Note how the data is similar in shape but overall much more scattered than in Figure \ref{fig:lengthscales}, which uses the definition of (\ref{eq:levapp}) instead. Also shown are the re-calibrated Zahn scale (this time with $(J\Pe)_c = 2$) and corresponding viscous scale $l_\nu$.  }
\label{fig:appendix}
\end{figure}

The values of $l_e$ measured using (\ref{eq:lew}) are shown in Figure \ref{fig:appendix}, while those using (\ref{eq:levapp}) are shown in the main text in Figure \ref{fig:lengthscales}. In both cases the eddy scale typically decreases with increasing $J \Pe_L$, and is reasonably independent of the Reynolds number. We can clearly see however that the data shown in Figure  \ref{fig:appendix} has much more scatter than in Figure \ref{fig:lengthscales}, a result that can be attributed to the relatively poor conditioning of equation (\ref{eq:lew}). Indeed, since $a_w(l)$ has a relatively flat tail, $l_e$ is very sensitive to small variations in $a_w(l)$. For this reason, we ultimately chose to use the definition of $l_e$ based on $a_v(l)$ in this work (see equation \ref{eq:levapp}), which does not suffer from this poor conditioning.

%\bibliographystyle{apj}
%\bibliography{NSF-bib}

\providecommand{\noopsort}[1]{}\providecommand{\singleletter}[1]{#1}%

\end{document}